\documentclass[a4paper,11pt]{article}
\pdfoutput=1 

\usepackage{jheppub} 

\usepackage[T1]{fontenc} 
\usepackage{booktabs}

\def\eps{\epsilon}

\newcommand \vev [1] {\langle{#1}\rangle}
 
\newcommand{\p}[1]{(\ref{#1})}

\newcommand{\cN}{{\cal N}}

\newcommand{\pa}{\partial}
\newcommand{\ep}{\epsilon}

\renewcommand{\a}{\alpha}

\newcommand{\la}{\lambda}

\newcommand{\da}{{\dot\alpha}}
\newcommand{\db}{{\dot\beta}}

\newcommand{\tr}{\mbox{tr}}

\title{Pentagon Wilson loop with Lagrangian insertion at two loops in ${\mathcal N}=4$ super Yang-Mills theory}
\author[a]{Dmitry Chicherin}
\author[b]{Johannes Henn}

\affiliation[a]{LAPTh, Universit\'{e} Savoie Mont Blanc, CNRS, B.P. 110, F-74941 Annecy-le-Vieux, France}
\affiliation[b]{Max-Planck-Institut f\"{u}r Physik, Werner-Heisenberg-Institut, D-80805 M\"{u}nchen, Germany}

\emailAdd{chicherin@lapth.cnrs.fr}
\emailAdd{henn@mpp.mpg.de}

\preprint{LAPTH-013/22, MPP-2022-37}

\abstract{
We compute the two-loop result for the {null pentagonal} Wilson loop with a Lagrangian insertion (normalized by the Wilson loop without insertion) in {planar}, maximally supersymmetric Yang-Mills theory. This finite observable is closely related to the Amplituhedron, and it is reminiscent of finite parts of planar two-loop five-particle scattering amplitudes. We verify that, up to this loop order, the leading singularities are given by the same conformally invariant expressions that appear in all-plus {pure Yang-Mills} amplitudes. The accompanying weight-four transcendental functions are expressed in terms of the pentagon functions space known from planar two-loop five-particle amplitudes, but interestingly only a subset of the functions appears. Being a function of four dimensionless variables, the observable has interesting asymptotic limits. We verify that our analytic result is consistent with soft and collinear limits, and find an intriguingly simple pattern in the multi-Regge limit. Thanks to the new result we can also conjecturally predict, for general kinematics, the maximal weight piece of the planar three-loop five-particle all-plus amplitude in pure Yang-Mills theory. Motivated by the Amplituhedron geometry, we investigate positivity properties of the integrated answer. Generalizing previous results at four particles, we find numerical evidence that the two-loop five-particle result has uniform sign in a kinematic region suggested by the loop Amplituhedron.
}

\begin{document}

\setcounter{tocdepth}{2}
\maketitle
\setcounter{page}{1}


\section{Introduction}

The duality between scattering amplitudes and null polygonal Wilson loops in maximally supersymmetric Yang-Mills theory (${{\mathcal N}=4}$ sYM) \cite{Alday:2007hr,Drummond:2007aua,Brandhuber:2007yx} has led to numerous advances in the area of scattering amplitudes.
For example, in ${{\mathcal N}=4}$ sYM the planar loop integrand is determined from powerful recursion relations \cite{ArkaniHamed:2010kv}, or, surprisingly, via a beautiful geometric construction \cite{Arkani-Hamed:2013jha}.
Scattering amplitudes with many external particles and to high loop orders can be bootstrapped based on educated guesses of the relevant function space, and many further insights \cite{Dixon:2011pw,Drummond:2014ffa,Caron-Huot:2016owq,Caron-Huot:2019bsq,Caron-Huot:2020bkp}.
At the same time, the novel insights into the relationship between certain loop integrands and transcendental functions \cite{ArkaniHamed:2010gh,Henn:2013pwa,Henn:2020omi}, and tools for handling these functions \cite{Goncharov:2010jf,Duhr:2019tlz}, have had a crucial impact on computing loop integrals in QCD that are relevant for phenomenology.

However, there are two issues where further progress would be very important. Firstly, despite the novel insights on loop integrands, only in few cases it is actually possible to perform the integrations directly. (See \cite{Herrmann:2019upk} for some ideas.) One important technical difficulty is the fact that although the remainder and ratio functions \cite{Drummond:2008vq} are infrared and ultraviolet finite, it is not known how to write them in a manifestly finite way. A second issue is that due to dual conformal symmetry \cite{Drummond:2007au}, non-trivial amplitudes that can be studied in ${{\mathcal N}=4}$ sYM have at least $n\ge 6$ particles, and they depend on $(3n-15)$ dimensionless kinematic variables. This is to be compared with amplitudes in QCD, which depend on $(3n-11)$ kinematic variables, and where already the cases $n=4,5$ are of enormous current interest, see for example \cite{Czakon:2021mjy,Caola:2021izf} and references therein. It would therefore be desirable to have a relevant observable in ${{\mathcal N}=4}$ sYM that, for the same number of particles, has similar variable dependence as QCD amplitudes, and ideally could be computed in four dimensions, to fully exploit the novel integrand insights.

Perhaps surprisingly, such an observable exists, and it is closely related to the Wilson loops and scattering amplitudes usually studied.
Consider an $n$-cusp null polygonal Wilson loop with a Lagrangian insertion, normalized by the Wilson loop without insertion \cite{Alday:2011ga},
\begin{align}
F_n = {\pi^2}\frac{\vev{W_n {\cal L}(x_0)}}{\vev{W_n}}\,. \label{ratioFn}
\end{align}
This object has a number of desirable features. 
All cusp divergences of the Wilson loops cancel in the ratio, and its dual conformal symmetry is realized without anomalies. 
As a result, the $n$-point observable is a function of $(3n-11)$ dimensionless variables, just like generic QFT amplitudes. 
Furthermore, one can fix a dual conformal frame where the insertion point is sent to infinity, in which case the kinematic variables become identical to those for $n$-particle scattering amplitudes.
This means that already at four particles, one obtains a non-trivial function of one variable, $z=t/s$.
The latter has been computed perturbatively at weak \cite{Alday:2011ga,Alday:2012hy,Alday:2013ip}, and at strong coupling \cite{Alday:2011ga}.

The new observable is closely related to the Wilson loop / scattering amplitude. Consider differentiating $\log \vev{W_n}$ in the coupling. In this way, one obtains precisely the ratio \p{ratioFn} integrated over the Lagrangian insertion point $x_0$. In this sense the new observable can be thought of as an integrand of the (logarithm of the) Wilson loop / scattering amplitude, but where all except the final $x_0$ integrations have been carried out. It is the final integration over $x_0$ that leads to the divergences in the Wilson loop / scattering amplitude. This fact has recently been used to compute the four-loop non-planar cusp anomalous dimension \cite{Henn:2019swt}.

What structures does one find in the perturbative expansion of this new quantity $F_n$? 
As we will see, thanks to its definition, it inherits many properties of scattering amplitudes. 
For example, the $L$-loop contribution is expected to be a sum over a set of leading singularities,
multiplied by transcendental functions of uniform weight $2L$.
Just like for scattering amplitudes, there are very natural questions that we can ask:
What set of leading singularities can appear? What is the space of transcendental functions that can appear?

In a recent paper \cite{Chicherin:2022bov}, the present authors studied the structural and symmetry properties of the $n$-point observable, and gave first answers to these questions.
Building upon formulas and intuition from scattering amplitudes, we found a simple form of the leading singularities of $F_n$ in terms of a Grassmannian integral.
This integral just produces one leading singularity at four points, which can be chosen as an overall normalization factor.
The first non-trivial case is then five points, where we find six linearly independent leading singularities.
Amazingly, the leading singularities are the same (once multiplied with an overall Parke-Taylor factor) as those observed to appear in all-plus helicity amplitudes in QCD \cite{Badger:2019djh,Henn:2019mvc}, and they share the same conformal symmetry \cite{Chicherin:2022bov}. This is not at all obvious, but can be shown for the Grassmannian formula given in \cite{Chicherin:2022bov}. Whether all leading singularities are given by 
this formula is an open question.

In the same paper \cite{Chicherin:2022bov}, the $F_n$ was evaluated at tree-level and one-loop, providing evidence for the proposed leading singularity formula.
Moreover, the one-loop result gives first examples of the types of transcendental functions one can expect. 
It was found that the answer can be written in terms of finite one-loop integrals, and hence is reminiscent of finite parts of one-loop $n$-particle scattering amplitudes.
What is more, a very curious relationship to the pure Yang-Mills all-plus amplitude was found: $F_n$ at $L$ loops is equal, up to an overall normalization factors, to the leading transcendental terms of the $(L+1)$-loop 
all-plus scattering amplitudes. This was shown by explicit computation at $L=0$ and $L=1$ for any $n$, and at $L=2$ for $n=4$.

Thanks to the connection to scattering amplitudes, one can also make progress on the second question raised above, namely which transcendental functions can appear. The all-loop integrand of planar ${{\mathcal N}=4}$ sYM \cite{ArkaniHamed:2010kv,Arkani-Hamed:2013jha} can be beautifully defined geometrically in terms of the Amplituhedron \cite{Arkani-Hamed:2013jha}. Very recently, it was shown that a generalization of the Amplituhedron idea allows one to express the new observable in terms of a sum of integrals associated to certain negative geometries. The latter define a novel type of finite, four-dimensional loop integrals \cite{Arkani-Hamed:2021iya}. These integrals were analyzed so far at four points, where certain infinite classes of integrals were computed.

The Amplituhedron story also motivates to look for positivity properties of scattering amplitudes \cite{Arkani-Hamed:2014dca,Dixon:2016apl}.
Indeed, in \cite{Arkani-Hamed:2021iya} it was found that the four-point results have definite sign properties.
It would be very interesting to study this property for higher-point amplitudes.

Motivated by these new developments and findings, in this paper we perform a perturbative two-loop calculation of the null pentagonal Wilson loop with Lagrangian insertion. 
In this way we can test the observations of \cite{Chicherin:2022bov} regarding leading singularities at two loops, which is highly non-trivial.
Moreover, this is the first time that this quantity is evaluated beyond one loop beyond four points.
The resulting transcendental functions are expected to be of weight four and depend on four independent kinematic variables, which is very interesting.
For example, we test the behaviour of the function in soft and collinear limits, and study its multi-Regge limit. 
We also test positivity properties of the integrated answer.

This work is organized as follows. In Section \ref{sec:def}, we recall the definition of the finite observable \p{ratioFn}, as well as perturbative results in the four-particle case, and their key properties. In Section \ref{sec:5pt}, we discuss the five-particle observable. We present its leading singularities and recall definitions of the pentagon functions. The latter turn out to be the transcendental functions which describe loop corrections of the five-particle observable. At the end of the Section we present our main result, the five-particle observable in the two-loop approximation. 
In Section \ref{sec:calculationandresult}, we outline the two-loop calculation. We construct the integrands from the five-particle MHV amplitude integrands available in the literature, and then perform loop integrations. We also calculate soft, collinear and multi-Regge limits of the five-particle observable. In Section \ref{sec:positivity}, we discuss positivity properties of the five-particle perturbative corrections.
In Section \ref{sect:all-plus}, we review the conjectured duality between the finite observable of Section \ref{sec:5pt} and the five-particle all-plus amplitude in pure Yang-Mills theory. Based on our results, we predict the maximally transcendental part of the planar three-loop all-plus amplitude in pure Yang-Mills. 
We mention directions for future research in Section \ref{sec:outlook}. 
In the Appendix, we recall definitions of the planar pentagon functions and their basic properties.

\section{The $n$-cusp Wilson loop with Lagrangian insertion}
\label{sec:def}

\subsection{Definition and symmetry properties}

Let us define the ratio of the Wilson loop correlation function with the Lagrangian of ${\cal N}=4$ sYM and the vacuum expectation value of the Wilson loop in the fundamental representation of the gauge group $SU(N_c)$, 
\begin{align}
\frac{1}{\pi^2} F_n(x_1,\ldots,x_n;x_0) = \frac{\vev{W_F[x_1,\ldots,x_n]\, {\cal L}(x_0)}}{\vev{W_F[x_1,\ldots,x_n]}}  \,, \label{FnBos}
\end{align}
following the conventions of ref.  \cite{Chicherin:2022bov}. (For $n=4$ points, a slightly different normalization was used in ref. \cite{Arkani-Hamed:2021iya}.)
The Wilson loop contour $[x_1,\ldots,x_n]$ is an $n$-cusp polygon with light-like edges, that is 
\begin{align}
(x_{i+1}-x_i)^2 = 0,\qquad i=1,\ldots,n,
\end{align}
embedded in Minkowski space, $x_{n+1}\equiv x_1$. In the following we work in the planar limit $N_c \to \infty$. The Lagrangian in \p{FnBos} is the so-called {\em chiral and on-shell} form of the Lagrangian ${\cal L}$, which coincides with the standard Lagrangian of ${\cal N}=4$ sYM modulo total derivatives and equations of motion (see definition in eq. (A.13) of \cite{Eden:2011yp}). This choice is motivated by the correlator/amplitude duality \cite{Eden:2010ce,Eden:2010zz,Eden:2011yp} (see \cite{Chicherin:2022bov} for a discussion). The ratio $F_n$ contains parity-odd contributions at $n >4$ because of chirality of the Lagrangian insertion.

Being a finite quantity in $d=4$ dimensions, the ratio $F_n$ in eq. (\ref{FnBos}) has a dual conformal symmetry, i.e. a conformal symmetry acting on the $x_{i}$ variables, with dual conformal weight zero at points $x_i$, $i\in \{1,\ldots n\}$, and dual conformal weight $+4$ at the Lagrangian insertion point $x_0$.
The symmetry implies that $F_n$ can be written as a factor carrying the dual conformal weights, multiplied by a function of $(3n-11)$ dual-conformal cross-ratios.

The perturbative expansion of $F_n$ in the coupling $g^2= g_{\rm YM}^2 N_{c}/(16 \pi^2)$ is as follows
\begin{align}
F_n = \sum_{L \geq 0} (g^2)^{1+L} F_n^{(L)} \,. \label{FnPertSer}
\end{align}
In the following we often refer to $F^{(0)}_n$ as the Born-level observable, and $F^{(L)}_n$ with $L>0$ as the $L$-loop observable. 
The expected uniform transcendental weight property of $\cN =4$ sYM theory suggests that the 
terms of the perturbative expansion should be of the following form
\begin{align}\label{eq:LSandtranscendental}
F_n^{(L)} = \sum_{s} R_{n,s}\, g^{(L)}_{n,s} \,.
\end{align}
Here $\{ R_{n,s} \}$ is a set of leading singularities labeled with an index $s$, which are rational functions of $x_1,\ldots,x_n,x_0$,
and the accompanying $g^{(L)}_{n,s}$ are pure functions of transcendental weight $2 L$. In particular, at the Born-level ($L =0$) we have $g^{(0)}_{n,s} = 1$, and $F_n^{(0)}$ is a rational function. The leading singularities carry a dual conformal weight $+4$ at $x_0$, and $g_{n,s}^{(L)}$ are functions of $(3n-11)$ cross-ratios. Conjecturally, all planar leading singularities $R_{n,s}$ at any $n$ are known explicitly \cite{Chicherin:2022bov}.

We saw that $F_n$ has a dual conformal symmetry in $x$-coordinates. We can use dual-conformal transformations, without loss of generality, to send $x_0$ to space-time infinity,
\begin{align}
f_n (x_1,\ldots,x_n) := \lim_{x_0 \to \infty} (x_0^2)^4 \, F_n(x_1,\ldots,x_n;x_0) \,. 
\label{fnLimFn}
\end{align} 
This amounts to fixing a dual-conformal frame. 
This object contains the same amount of information as $F_n$, and the dependence on $x_0$ can be easily restored for a given $f_n$. 
We also define the leading singularities of $F_n$ in this frame,
\begin{align}
r_{n,s}(x_1,\ldots,x_n) := \lim_{x_0 \to \infty} (x_0^2)^4  R_{n,s}(x_1,\ldots,x_n;x_0) \,. \label{rLimR}
\end{align}
In other words, $\{ r_{n,s} \}$ are leading singularities of $f_n$. The pure functions $g^{(L)}_{n,s}$ from eq.~\p{eq:LSandtranscendental} keep their form in the frame $x_0 \to \infty$. 

\begin{figure}
    \centering
    \includegraphics[height=4.0cm]{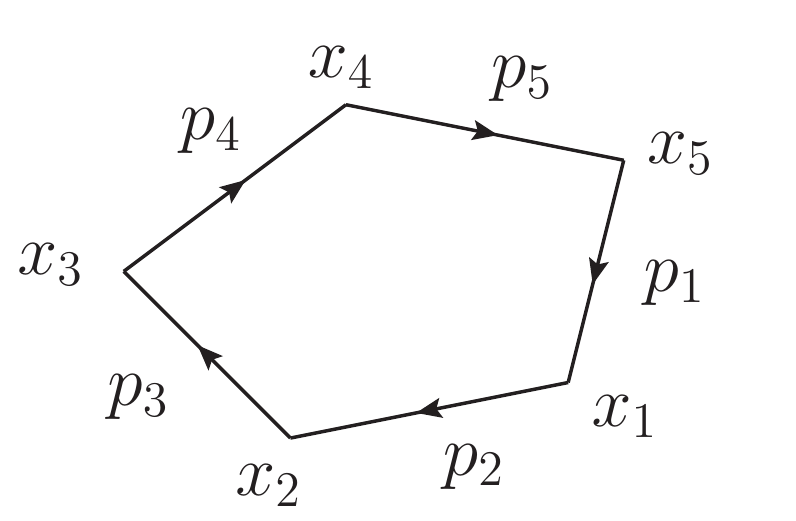}
    \caption{Momenta and dual-momenta assignment to the five-cusp light-like Wilson loop contour.}
    \label{fig:5ptcontour}
\end{figure}

The frame $x_0 \to \infty$ has the advantage that it is easy to connect the kinematics to familiar momentum space terminology.
This is seen as follows. In this frame, the leading singularities are Poincar\'{e}-invariant functions of $n$ points separated by light-like distances, i.e. they are invariant under space-time shifts 
and Lorentz transformations. Equivalently, we can say that they are functions of $n$ light-like momenta ($p_i^2 = 0$)
\begin{align}
p_{i} = x_{i} - x_{i-1}  \label{xp}
\end{align} 
(with $x_{n+1} \equiv x_{1}$)
which satisfy momentum conservation (see Fig.~\ref{fig:5ptcontour})
\begin{align}
\sum_{i=1}^{n} p_i = 0 \,.    
\end{align}
As usual, it is convenient to introduce helicity spinor parametrization of the light-like momenta, $(p_i)^{\da\a} \equiv p^\mu_i \sigma_\mu^{\da\a} = \la_i^{\a} \tilde\la^{\da}_i$. Since we have already fixed dual conformal symmetry, this is exactly the same kinematics as for a non-dual-conformal scattering amplitude of $n$ massless particles. Indeed, we will see that the known results bear close resemblance to scattering amplitudes in non-supersymmetric Yang-Mills theory. In the following we adopt the amplitude terminology and refer to $f_n$ as $n$-particle. 

In ref. \cite{Chicherin:2022bov} it was found that $F_n$ has a hidden conformal symmetry (and a Yangian-like symmetry). The latter is best seen in the frame $x_0 \to \infty$, and when multiplying $f_n$ with the Parke-Taylor factors
\begin{align}\label{eq:ParkeTaylor}
{\rm PT}_{n} := \frac{1}{\vev{12} \vev{23} \ldots \vev{(n-1) n} \vev{n1}} \,,
\end{align}
where $\vev{ij} := \la^{\alpha}_i \la_{j \alpha}$ (later on, we will also need the conjugated spinor bracket $[i j] := \tilde\la_{i\, \dot\alpha} \tilde\la_{j}^{ \dot\alpha}$).
It was found \cite{Chicherin:2022bov} that the leading singularities of $f_n$ are conformally invariant, namely
\begin{align}\label{eq:hiddenconformalinvariance}
\mathbb{K}_{\a\da}  \left(  {\rm PT}_{n} r_{n,s} \right) = 0\,,
\end{align}
with the conformal boost generator in momentum space of $n$-particle scattering \cite{Witten:2003nn}
\begin{align}
\mathbb{K}_{\a\da} =  \sum_{i=1}^n \frac{\pa^2}{\pa \la_{i}^{\alpha} \pa \tilde\la_{i}^{\dot\alpha}} \,. \label{KboostDef}
\end{align}

\subsection{Review of perturbative results in the four-particle case} \label{f4pt}
Let us illustrate these definitions for the known four-particle results. These will also be useful in the following, as they serve as important consistency checks for our novel five-particle results.
Dual conformal symmetry tells us that we can write
\begin{align}
F_4 = \frac{x_{13}^2 x_{24}^2}{x_{01}^2 x_{02}^2 x_{03}^2 x_{04}^2} G_4(g;z) \,, \label{F4G4}
\end{align}
where $x_{ab}^2 := (x_a - x_b)^2$ and $z$ is the unique dual-conformal cross-ratio at $n=4$, 
\begin{align}
z= \frac{x_{13}^2 x_{02}^2 x_{04}^2}{x_{24}^2 x_{01}^2 x_{03}^2} \,.
\end{align}
Equivalently, in the frame $x_0\to \infty$, 
we write the dual-conformal cross-ratio 
(in momentum-space notation) as
\begin{align}
z= \frac{t}{s} \,
\end{align}
in terms of the Mandelstam variables specifying the four-particle massless kinematics, 
\begin{align}
s = (p_1+  p_2)^2 \;,\quad 
t= (p_2 + p_3)^2 \,.  \label{st4pt}  
\end{align}
There are several kinematic regions of the four-particle scattering. We will be primarily concerned with the Euclidean region: $s < 0$ and $t < 0$. In this region $z>0$.

The function $G_4$ has been computed to three loops in perturbation theory \cite{Alday:2012hy,Alday:2013ip,Henn:2019swt}. Here we reproduce the first three terms of the expansion,
writing
\begin{align}
G_4 = \sum_{L \geq 0} (g^2)^{1+L} G_4^{(L)} \,. \label{GPertSer}
\end{align}
The explicit polylogarithmic expressions are as follows,
\begin{align}
G^{(0)}_{4}(z)=& -1 \,,\label{Fweak0} \\
G^{(1)}_{4}(z)=& \log^2 z + \pi^2 \,, \label{Fweak1}\\
G^{(2)}_{4}(z)=& -\frac{1}{2} \log^4 z  
+ \log^2 z \left[ \frac{2}{3} \text{Li}_2\left(\frac{1}{z+1}\right)+\frac{2}{3} \text{Li}_2\left(\frac{z}{z+1}\right)-\frac{19 \pi ^2}{9} \right]  \label{Fweak2}\\
&\hspace{-1cm} + \log z \left[4 \text{Li}_3\left(\frac{1}{z+1}\right)-4 \text{Li}_3\left(\frac{z}{z+1}\right) \right] 
+ \frac{2}{3} \left[\text{Li}_2\left(\frac{1}{z+1}\right)+\text{Li}_2\left(\frac{z}{z+1}\right)-\frac{\pi ^2}{6}\right]^2 \nonumber \\
&\hspace{-1cm} +\frac{8}{3} \pi ^2 \left[\text{Li}_2\left(\frac{1}{z+1}\right)+\text{Li}_2\left(\frac{z}{z+1}\right)-\frac{\pi ^2}{6}\right]
+ 8 \text{Li}_4\left(\frac{1}{z+1}\right)+8 \text{Li}_4\left(\frac{z}{z+1}\right)-\frac{23 \pi ^4}{18}\,. \nonumber
\end{align}
Each $G_4^{(L)}$ is a pure function of uniform transcendental degree $2L$. The provided polylogarithmic expressions are single-valued in the Euclidean region $z > 0$. 

The Amplituhedron construction provides a natural way to define the planar integrand of $F_4$ directly from geometric equations \cite{Arkani-Hamed:2021iya}. In this approach, the loop corrections \p{Fweak0}-\p{Fweak2} are obtained as sums of finite four-dimensional Feynman integrals corresponding to negative geometries.

The Amplituhedron also suggests to look for positivity properties of integrated functions \cite{Arkani-Hamed:2014dca,Dixon:2016apl}.
The four-point one-loop Amplituhedron corresponds to the kinematic region $z>0$, so it is interesting to investigate the sign properties of $f_{4}^{(L)}$ in that region.
Indeed, it was found in ref. \cite{Arkani-Hamed:2021iya} that the available four-point loop corrections are either only positive or only negative in the Euclidean region $z>0$, and the sign alternates with the loop order,
\begin{align}
f_{4}^{(0)}(z)\bigr|_{z>0} < 0 \,,\quad
f_{4}^{(1)}(z)\bigr|_{z>0} > 0 \,,\quad
f_{4}^{(2)}(z)\bigr|_{z>0} < 0 \,,\quad \ldots \label{f4ptPos}
\end{align}
The available perturbative data suggests that $f_4^{(L)} = s t G_4^{(L)}$ is positive within the Euclidean region at odd $L$, and it is negative at even $L$.

Comparing to eq. \p{eq:LSandtranscendental}, we see that in the four-point case, there is only one leading singularity, namely 
\begin{align}
R_{4,0} = \frac{{x_{13}^2 x_{24}^2}}{{x_{01}^2 x_{02}^2 x_{03}^2 x_{04}^2}},
\end{align}
and $g_{4,0} = G_{4}$ in that formula.
It is instructive to present the leading singularity in momentum space and to discover its hidden conformal properties.
Employing eq. (\ref{rLimR}) and passing to momentum-space notation, we have
\begin{align}
r_{4,0} =  s\, t\,.
\end{align}
A quick calculation shows that after normalizing by the Parke-Taylor factor \p{eq:ParkeTaylor},
\begin{align}
{\rm PT}_{4} \, r_{4,0} = -\frac{[12]^2}{\vev{34}^2} \,. \label{PT4r40}
\end{align}
In this form, the conformal invariance of eq. (\ref{eq:hiddenconformalinvariance}) is manifest, since the conformal boost operator in eq. (\ref{KboostDef}) annihilates any expression that depends on $\lambda_i$ and $\tilde{\lambda}_j$ with $i\neq j$ only. 

This conformal invariance is related to another surprising property of $F_n$, also discovered in ref. \cite{Chicherin:2022bov}, namely a duality to all-plus amplitudes in the pure Yang-Mills theory. In particular, as discussed in Sect.~\ref{sect:all-plus}, 
\begin{align}
A^{(1)}_{{\rm YM},4}= {\rm PT}_{4} \, r_{4,0}    
\end{align} 
is the four-particle one-loop all-plus amplitude in the pure Yang-Mills theory.

In summary, based on the known perturbative results we can say that the (planar) four-point observable $F_{4}$ has a number of remarkable properties, namely: 
\begin{itemize}
\item $F^{(L)}_4$ has uniform transcendental weight $2L$; 
\item positivity in the kinematic region defined by the Amplituhedron kinematics;
\item hidden conformal symmetry of leading singularities;
\item duality to four-point all-plus amplitude in pure Yang-Mills theory.
\end{itemize}
The main goal of the present paper is to investigate to which extent these properties generalize to the five-point observable $F_{5}$. 
In order to do so we compute, for the first time, the two-loop corrections to $F_{5}$.

\section{Five-particle observable and novel two-loop result}
\label{sec:5pt}

The five-particle case is much richer than the four-particle one. Indeed, $F_5$ involves four independent dual conformal cross-ratios. Consequently, the loop corrections involve multivariable transcendental functions. We provide evidence that they are the pentagon functions, which describe five-particle massless scattering amplitudes in nonsupersymmetric gauge theories. 

\subsection{Five-particle kinematics}
\label{sec:5partkin}

As in the previous section, we find it convenient to work in the gauge $x_0 \to \infty$.
Let us introduce notations for the five-particle kinematics. 
In order to specify the kinematic configuration, we choose five adjacent two-particle Mandelstam variables
\begin{align}
{\bf s} := \{ s_{12} \,, \; s_{23} \,, \; s_{34} \,, \; s_{45} \,, \; s_{15} \} \,,\qquad s_{ij} := (p_i + p_j)^2\,, \label{sadj}
\end{align}
as well as the parity-odd Lorentz invariant
\begin{align}
\ep_5 := 4 \textup{i} \ep_{\mu\nu\rho\sigma} p_1^\mu p_2^\nu p_3^\rho p_4^\sigma \,. \label{e5}
\end{align}
The latter is not fully independent of the Mandelstam variables \p{sadj}. Indeed, $\Delta := (\ep_5)^2$ is a quartic polynomial in $s_{ij}$ given by the Gram determinant of four independent momenta, $\Delta= \det \bigl( s_{ij}|^{4}_{i,j=1} \bigr)$. Nevertheless, the parity-odd $\ep_5$ helps to distinguish kinematic configurations related by the parity conjugation, $\ep_5 \to - \ep_5$. 

Our choice of the kinematic variables reflects the discrete symmetries of $F_5$, or equivalently of $f_5$. Indeed, the geometry of the Wilson loop contour in \p{FnBos} is invariant under the cyclic shift $x_i \to x_{i+1}$ and the reflection $x_i \to x_{6-i}$. In momentum space notations, the first is given by  the cyclic shift of momenta labels,
\begin{align}
\tau( p_i ) := p_{i+1} \,, \label{Dihed1}
\end{align}
with $p_{6} \equiv p_{1}$, and
the latter is equivalent to the reflection of momenta labels, i.e.
\begin{align}
\rho( p_i ) := p_{6-i} \,, \quad i=1,\ldots,5\,. \label{Dihed2}
\end{align}
These discrete transformations $\tau$ and $\rho$ form the dihedral group 
\begin{align}
\tau^5 = \rho^2 = (\tau \rho)^2 = (\rho\tau)^2= \mathbf{1} \,.    
\end{align}
Let us note that $\ep_5$ is invariant under the dihedral transformations, $\tau(\ep_5) = \rho(\ep_5) = \ep_5$.

Similarly to the four-particle case, we specify the Euclidean kinematic region by requiring that all adjacent Mandelstam invariants are negative, 
\begin{align}
{\bf s } < 0 \; : \quad s_{12} < 0 \,, \; s_{23} < 0 \,, \; s_{34} < 0 \,, \; s_{45} <0 \,, \; s_{15} < 0 \,.   \label{Eucl5}
\end{align}
These inequalities implicate $(\ep_5)^2 = \Delta > 0$. 
Recalling definition \p{e5} of $\ep_5$, we conclude that the light-like momenta $p_i^\mu$ are complex valued in the Euclidean region, but their scalar products $s_{ij}$ are real-valued. The Euclidean region is invariant under the dihedral transformations \p{Dihed1}, \p{Dihed2}.

\subsection{Five-particle Amplituhedron geometry and kinematic region}
\label{sec:5partkinAmplituhedron}

We will be interested in testing positivity properties of the five-particle loop corrections $f_5^{(L)}$.
The question is, in what kinematic region should one expect special sign properties of the answer?
As the main motivation is the Amplituhedron geometry, it makes sense to see what it implies for the five-particle case.

The relevant geometry is the five-particle one-loop MHV amplituhedron, which is carved out by the following inequalities on the four-brackets  $\vev{abcd}=\det(Z_a,Z_b,Z_c,Z_d)$ of the momentum twistors \cite{Arkani-Hamed:2017vfh,Herrmann:2020qlt},
\begin{align}
& \vev{ijkl} > 0 \,, \qquad \text{for} \quad  \leq i<j<k<l \leq 5 \,, \label{ah1}\\
& \vev{ABii+1} > 0 \,, \label{ah2}\\
& \{ \vev{AB12},\vev{AB13},\vev{AB14},\vev{AB15} \} \qquad \text{has two sign flips.} \label{ah3}
\end{align}
Here the momentum twistors $Z_i$, $i=1,\ldots,5$, describe the external five-particle kinematics, and the twistor line $(AB)$ represents the loop integration. We take $(AB)$ to be the infinity bi-twistor, $(AB) = \ep^{\da\db}$, as this corresponds to the frame $x_0 \to \infty$ in space-time notations. Then we have $\vev{AB i i+1} = \vev{i i+1}$ and $\vev{ii+1jj+1} = \vev{ii+1}\vev{jj+1} x_{ij}^2$.
Thus, eqs. \p{ah1} and \p{ah2} imply that 
\begin{align}
s_{12} > 0 \,,\; s_{23} > 0 \,,\; s_{34} > 0 \,,\; s_{45} > 0 \,,\; s_{15} > 0 \,.
\end{align}
This is equivalent to the Euclidean region \p{Eucl5} up to an (irrelevant) overall sign flip of the Mandelstam variables. 

Moreover, eqs. \p{ah2} and \p{ah3} imply that the signs of $\{ \vev{12}, \vev{13}, \vev{14}, \vev{15}\}$ allowed by the Amplituhedron geometry are $\{ +,-,-,+\}$, $\{ +,+,-,+\}$ and $\{ +,-,+,+\}$. The configuration $\{ +,+,+,+ \}$ belongs to the Euclidean region \p{Eucl5} but it is forbidden by \p{ah3}. 
We find that $\ep_5>0$ on the configurations $\{ +,-,-,+\}$, $\{ +,+,-,+\}$, $\{ +,-,+,+\}$ and $\ep_5<0$ on the configuration $\{ +,+,+,+ \}$. 
Thus, in summary, the $\ep_5>0$ part of the five-particle Euclidean region \p{Eucl5} is the one-loop MHV Amplituhedron geometry (up an irrelevant overall sign flip).
This is consistent with a related discussion in \cite{Gehrmann:2015bfy}.
Therefore it will be particularly interesting to look for positivity properties of $F_5$ in that region.

\subsection{Conjecture for leading singularities}
\label{LS5pt}

In the $n$-particle case with $n > 4$, there are several leading singularities of $F_n$. The leading singularities of $F_n$ were analyzed in \cite{Chicherin:2022bov} in the planar limit, where it was conjectured that they are given by certain four-point and five-point momentum twistor functions. We denote them by $b_{ijkl}$, $b_{ijklm}$ in the frame $x_0 \to \infty$. It turns out that at $n=5$, one can write down ten such terms, six of which are linearly independent.
After normalizing by the Parke-Taylor prefactor ${\rm PT}_5$ of eq. \p{eq:ParkeTaylor}, they take the following simple form in the frame $x_0 \to \infty$ in spinor helicity notation,
\begin{align}
& {\rm PT}_5\, b_{1234} = -\frac{[23]^2}{\vev{45}\vev{51}\vev{14}} \;,\qquad 
{\rm PT}_5\, b_{2345} = -\frac{[34]^2}{\vev{51}\vev{12}\vev{25}} \,, \notag\\
& {\rm PT}_5 \, b_{1345} = -\frac{[45]^2}{\vev{12}\vev{23}\vev{31}} \;,\qquad  {\rm PT}_5 \, b_{1245}= -\frac{[51]^2}{\vev{23}\vev{34}\vev{42}} \,, \notag\\
& {\rm PT}_5 \, b_{1235} = -\frac{[12]^2}{\vev{34}\vev{45}\vev{53}}  \;,\qquad
{\rm PT}_5 \, b_{12345} = \frac{[13]^2}{\vev{24}\vev{45}\vev{52}} \,. \label{5ptrres}
\end{align}
Just as in the four-particle case, it is obvious in this form that the conformal invariance of eq. (\ref{eq:hiddenconformalinvariance}) holds. 
Conjecturally, the six functions from \p{5ptrres} are a basis for the leading singularities of the five-particle observable $f_5$ to all loop orders. 
In the present paper, we check to two loops, by the explicit calculation of $f_5^{(2)}$, that this is indeed the case.

We find it convenient to choose the following basis of six leading singularities,
\begin{align}
r_{5,0} & = b_{1245} + b_{2345} - b_{12345} \,, & \quad 
r_{5,1} & =  b_{2345}\,, & \quad r_{5,2} =  b_{1345}\,, \notag\\
r_{5,3} & =  b_{1245} \,, &
\quad r_{5,4} & =  b_{1235}\,, & 
\quad r_{5,5} =  b_{1234}\,. \label{f50}
\end{align}
One reason for this choice is that in this way, $f_5 = r_{5,0}$ at Born-level is among the leading singularities. 
The latter expression can be also written in a more symmetric form, avoiding spurious poles of individual $b$'s in expression \p{f50} for $r_{5,0}$,
\begin{align}\label{eq:resultf5tree}
r_{5,0} = -\frac{1}{2} \left( s_{12} s_{23} + s_{23} s_{34} + s_{34} s_{45} + s_{45} s_{15} + s_{15} s_{12} + \ep_5  \right)\,.
\end{align}
We note that it possesses the expected dihedral symmetry (recall eqs. \p{Dihed1}, \p{Dihed2}),
\begin{align}
\tau(r_{5,0}) = \rho(r_{5,0}) = r_{5,0} \,. \label{r50sym} 
\end{align}

\subsection{Duality to all-plus pure Yang-Mills amplitudes}
\label{sec:5ptAllplus}

Similarly to the four-particle case, the five-particle $f_5$ is related to the planar five-particle all-plus pure Yang-Mills scattering amplitude. This will be discussed in more details in Sect.~\ref{sect:all-plus}. At the lowest level in the weak-coupling expansion, the statement is as follows. The Born-level $f_5$ \p{eq:resultf5tree} normalized with the Parke-Taylor factor coincides with the one-loop all-plus amplitude, 
\begin{align}
A^{(1)}_{{\rm YM},5}  = {\rm PT}_{5} \, f_{5}^{(0)} \,.    \label{A1YM5pt} 
\end{align}

It is also instructive to provide expressions for the remaining five leading singularities in terms of the Mandelstam variables. They are as follows 
\begin{align}
& r_{5,1} = \frac{s_{34}}{s_{25}} \tr_{-}(p_2 p_3 p_4 p_5) \,,\quad
r_{5,2} = \frac{s_{45}}{s_{13}} \tr_{-}(p_3 p_4 p_5 p_1) \,,\quad
r_{5,3} = \frac{s_{15}}{s_{24}} \tr_{-}(p_4 p_5 p_1 p_2)
\,, \notag\\
& r_{5,4} = \frac{s_{12}}{s_{35}} \tr_{-}(p_5 p_1 p_2 p_3)
\,, \quad
r_{5,5} = \frac{s_{23}}{s_{14}} \tr_{-}(p_1 p_2 p_3 p_4) \,, \label{r5i}
\end{align}
where the Lorentz invariant
\begin{align}\tr_{-}(p_i p_j p_k p_l) :=& \frac{1}{2}\tr\left(\left(1-\gamma_5\right) {p}_i  {p}_j  {p}_k  {p}_l\right) \nonumber\\
 = & \vev{ij}[jk]\vev{kl}[li] \,,
\end{align}
involves the parity-odd $\ep_5$. In this representation, we observe that the five leading singularities can become singular if the scalar products of nonadjacent momenta vanish. We also notice the five leading singularities are related by the cyclic shifts $\tau$ and are mapped among each other by reflection $\rho$, see eqs. \p{Dihed1}, \p{Dihed2},
\begin{align}
\tau(r_{5,i}) = r_{5,i+1} \,,\quad
\rho(r_{5,i}) = r_{5,6-i} \,, \label{r5isym}
\end{align}
with $r_{5,6} \equiv r_{5,1}$.

\subsection{Structure of loop corrections}

Next, let us discuss the kinematic dependence of the pure transcendental functions representing the loop corrections of $F_5$. In accordance with \p{eq:LSandtranscendental}, the $g_{5,s}$ with $s=0,1\ldots 5$ are functions of four independent dual conformal cross-ratios, which can be chosen as 
\begin{align}
{\bf u} = \left\{ \frac{x_{25}^2 x_{10}^2 x_{40}^2}{x_{14}^2 x_{20}^2 x_{50}^2} \,,\;
 \frac{x_{13}^2 x_{40}^2}{x_{14}^2  x_{30}^2} \,,\;
 \frac{x_{24}^2 x_{10}^2}{x_{14}^2 x_{20}^2} \,,\;
 \frac{x_{35}^2 x_{10}^2 x_{40}^2}{x_{14}^2 x_{30}^2 x_{50}^2} \right\} \,.
\end{align} 
In the dual conformal frame $x_0 \to \infty$, 
the cross-ratios become ratios of the Mandelstam variables \p{sadj},
\begin{align}\label{eq:uvarsmomentum}
{\bf u} =  \left\{ \frac{s_{12}}{s_{15}}  \,,\;
 \frac{s_{23}}{s_{15}}    \,,\;
 \frac{s_{34}}{s_{15}} \,,\;
\frac{s_{45}}{s_{15}}  \right\} \,.
\end{align}

The explicit perturbative calculation reveals that the five-particle $f_5$ has the expected form of the loop corrections,
\begin{align}
f_{5}^{(L)} =& f_5^{(0)}\, g^{(L)}_0 ({\bf u}) +  \sum_{i=1}^5 r_{5,i}\, g^{(L)}_i ({\bf u}) \,,\label{fL5}
\end{align}
where $g_s^{(L)}({\bf u})$ with $s=0,1,\ldots,5$ are pure functions of transcendental weight $2 L$ of the variables (\ref{eq:uvarsmomentum}); and $f_5^{(0)} = r_{5,0}$ \p{eq:resultf5tree}. The pure functions posses the same dihedral symmetries as the accompanying leading singularities \p{r50sym} and \p{r5isym}. Namely, $g_0^{(L)}$ is invariant under cyclic shifts and reflections
\begin{align}
\tau\left( g_0^{(L)} \right) = \rho\left( g_0^{(L)} \right)= g_0^{(L)}  \,.
\end{align}
whereas all $g_i^{(L)}$ are related by cyclic shifts,
\begin{align}
\tau\left( g_i^{(L)} \right) = g_{i+1}^{(L)} \,,\quad
\rho\left( g_i^{(L)} \right) = g_{6-i}^{(L)}\, \label{gisym}
\end{align}
with $ g_{6}^{(L)} \equiv  g_{1}^{(L)}$. 

Let us collect the previously known five-particle results up to one-loop order. We have already mentioned the Born-level ($L = 0$) answer $f_5^{(0)}$ \p{eq:resultf5tree}, which in notations \p{fL5} is as follows,
\begin{align}
g^{(0)}_0 = 1 \,, \quad   {\rm and}\quad   g^{(0)}_{i} = 0 \quad {\rm for} \quad i=1,\ldots,5 \,.
\end{align}
At one loop $L = 1$, we have~\cite{Chicherin:2022bov},
\begin{align}
g^{(1)}_0 =& \log\left( \frac{s_{15}}{s_{23}} \right) \log\left( \frac{s_{12}}{s_{34}} \right) - \frac{\pi^2}{6} + {\rm cyclic} \,, \label{g01loop}\\
g^{(1)}_1 =&- {\rm Li}_{2}\left(1- \frac{s_{12}}{s_{34}} \right) -  {\rm Li}_{2}\left(1- \frac{s_{15}}{s_{34}} \right) - \log\left(  \frac{s_{12}}{s_{34}}\right)   \log\left(  \frac{s_{15}}{s_{34}}\right)  + \frac{\pi^2}{6}    \,, \label{g1loop}
\end{align}
and the remaining $g^{(1)}_i$ are cyclic shifts of $g^{(1)}_1$, see \p{gisym}.

In this paper, we will compute $f_5$ to two loops $L =2$. It is in agreement with the leading singularities and the uniform transcendentality of \p{fL5}. Before proceeding to calculations, let us discuss the relevant space of transcendental functions that $g^{(2)}_{5,s}$ from \p{fL5} belong to.

\subsection{Pentagon functions}
\label{sect:pentfun}

We have already mentioned that the Wilson loop ratio in the frame $x_0 \to \infty$ is closely related to the all-plus YM scattering amplitude. Hence, in order to describe loop corrections of the planar five-particle $f_5$ we would need the same collection of transcendental functions as for massless planar five-particle scattering in non-supersymmetric gauge theories. They are the {\em planar pentagon functions} \cite{Gehrmann:2015bfy,Gehrmann:2018yef}. 

The pentagon functions have been actively studied over the past several years. The principal interest to them owes to two-loop calculation of QCD amplitudes. The latter are required to provide NNLO theoretical predictions for cross-sections measured in the ongoing collider experiments \cite{Chawdhry:2019bji}. The pentagon functions are an indispensable ingredient in these calculations. 

The planar pentagon functions have been thoroughly studied in \cite{Gehrmann:2018yef}. They have been classified up to the transcendental weight four that is required for undertaking two-loop amplitude calculations. This ref. also provides numerical routines for their evaluations within the Euclidean region \p{Eucl5} as well as in the physical scattering channels. More recently, the planar pentagon functions have been complemented with the nonplanar ones \cite{Chicherin:2020oor} which are required in the calculation of nonplanar corrections of the five-particle massless scattering amplitudes. In the present study, only planar pentagon functions will be required, and as discussed in section \ref{sec:5partkinAmplituhedron}, we are mostly interested in their evaluations in the Euclidean region \p{Eucl5}. Thus, we will rely on the planar results of ref. \cite{Gehrmann:2018yef}.

The pentagon functions are graded by their transcendental weight $w$. At weight $w>0$, there are several linearly independent pentagon functions $p^{(w)}_a$ which we label with an index $a$.\footnote{In ref.~\cite{Gehrmann:2018yef}, the pentagon functions are denoted as $f_{w,a}$ (and also $f_{w,a}^{(i)}$). We changed notations for the pentagon functions here to avoid confusion with the observable $f_n$ \p{fnLimFn}.} For the sake of convenience, we also consider the weight-zero case: $p^{(0)}=1$. Not only the pentagon functions are linearly independent, they are also algebraically independent. Namely, any nonzero polynomial in pentagon functions (of weight less of equal $w$) with constant coefficients does not vanish identically.

The polylogarithmic nature of the pentagon functions is revealed upon differentiating them in kinematic variables. Differentiation of $p^{(w+1)}_a$ with $w \geq 0$ results in a linear combination (or a polynomial) in the pentagon functions of lower weight,
\begin{align}
d p^{(w+1)}_a = \sum_{k,b} A^{k}_{a,b} \, p^{(w)}_b d\log(W_k)   + \text{terms with } p^{(w')} \text{ at } w' <w \,, \label{diffpent}
\end{align}
see also \p{DEpent1}--\p{DEpent4} for more details.
Here $A^{k}_{a,b}$ are rational numbers. Summation index $b$ runs over pentagon functions of weight $w$, while index $k$ runs over 26 letters of the planar pentagon alphabet \cite{Gehrmann:2015bfy}. We denote the alphabet letters as $\{ W_{k} \}_{k=1}^{20} \cup \{ W_k \}_{k=26}^{31}$ following notations of \cite{Chicherin:2017dob}.\footnote{Five letters $\{ W_k \}_{k=21}^{25}$ belong to the nonplanar extension of the pentagon alphabet. We do not use them in the present work.} We recall explicit expressions for the pentagon letters as well as more details on the pentagon functions in App.~\ref{App:pent}.

Iterating eq. \p{diffpent} one can obtain iterated integral representation for the pentagon functions. It is also useful to consider the symbol projection (see~\cite{Duhr:2019wtr} for a review) of the pentagon functions. The symbol of a pentagon functions is obtained by dropping out the transcendental constants in the iterated integral expression. Namely, the symbol of a weight-$w$ pentagon function is a linear combination with rational numbers of the length-$w$ words $[W_{i_1},\ldots,W_{i_w}]$ formed from the pentagon alphabet letters. Zero locus of the first entries are branch points of the pentagon functions. Thus, the allowed first entries of the pentagon function symbols are five adjacent Mandelstam variables \p{sadj}. In our calculations, we work with the iterated integrals, i.e. we keep 'beyond the symbol' terms.

In the five-particle setting the parity conjugation is nontrivial. The pentagon functions are defined to have definite parity, i.e. they are either even or odd upon $\ep_5 \to -\ep_5$.

\subsection{Novel five-particle two-loop result}
\label{sec:res2loop}

In this Section, we present the five-particle $f_5$ in the two-loop approximation. It is the result of an explicit two-loop calculation provided in the next Section. We find that it agrees with the expected form \p{fL5} of the leading singularities,
\begin{align}
f_{5}^{(2)} =& f_5^{(0)}\, g^{(2)}_0 ({\bf u}) +  \sum_{i=1}^5 r_{5,i}\, g^{(2)}_i ({\bf u}) \,.\label{fL5twoLoop}
\end{align}
The loop corrections $g^{(2)}_0$ and $g^{(2)}_i$ with $i=1,\ldots,5$ are pure functions of the transcendental weight four, which are polynomials in the planar pentagon functions of \cite{Gehrmann:2015bfy}. Namely, $g^{(2)}_0$ and $g^{(2)}_i$ are linear combinations with rational coefficients of the monomials 
\begin{align}
p^{(4)}_a ,\; p^{(3)}_a p^{(1)}_b,\; p^{(2)}_a p^{(2)}_b ,\; p^{(2)}_a p^{(1)}_b p^{(1)}_c ,\; p^{(1)}_a p^{(1)}_b p^{(1)}_c p^{(1)}_d ,\; \zeta_3 p^{(1)}_a ,\; \pi^2 p^{(2)}_a ,\; \pi^2 p^{(1)}_a p^{(1)}_b \label{g2inf}
\end{align}
and two weight-4 transcendental constants $\phi^{(4)}_0$ and $\phi^{(4)}_1$. 
Note that despite the fact that the planar pentagon functions \cite{Gehrmann:2018yef} are not dimensionless, they are organized in dimensionless expressions in $g^{(2)}_0$ and $g^{(2)}_i$.

The one-loop $g^{(1)}_0,\,g^{(1)}_i$ are parity even, while the two-loop functions involve parity-odd pentagon functions. Namely, $g^{(2)}_0$ is parity-even and $g^{(2)}_i$ contains both parity-odd and parity-even terms.

We collected the pentagon function expressions for the two-loop (as well as one-loop) $f_5$ in computer-readable ancillary files. This representation can be easily evaluated numerically using the public code provided with \cite{Gehrmann:2018yef}. We also provide iterated integral expressions (and consequently the symbol expressions) for one-loop and two-loop $f_5$ in the ancillary files. 

We provide numerical values of $f_5$, 
\begin{align}
f_5^{(0)} = -7.81495 \,,\quad
f_5^{(1)} =  96.3895 \,,\quad
f_5^{(2)} = -1398.9113 \,,
\end{align}
at the reference point (with $\ep_5 >0$)
\begin{align}
s_{12} = -1\,,\;
s_{23} = -\frac{31}{11}\,,\;
s_{34} = -\frac{59}{43}\,,\;
s_{45} = -\frac{23}{13}\,,\;
s_{15} = -\frac{47}{61}\,.
\end{align}

Let us now discuss some features of the result.
The iterated integral expression of $g_0^{(2)}$ involves 25 letters of the planar pentagon alphabet, namely the letter $W_{31}$ is absent. Similarly, planar pentagon letters $W_{6},W_{8},W_{10},W_{31}$ are absent in the iterated integral expression of $g_1^{(2)}$, so it involves 22 letters. The remaining functions $g^{(2)}_i$ ($i=2,\ldots,5$) are cyclic permutations of $g_1^{(2)}$, so they also involve 22 letters and the missing letters are the cyclic permutations of $W_{6},W_{8},W_{10},W_{31}$. Previously, it was noticed that the letter $W_{31}$ is absent from a properly defined finite hard part of the planar (as well as nonplanar) five-particle two-loop amplitudes in maximally supersymmetric Yang-Mills theory \cite{Abreu:2018aqd,Chicherin:2018yne} and supergravity \cite{Chicherin:2019xeg,Abreu:2019rpt}. The same is also true for the hard part of QCD amplitudes with gluons and quarks in the external states \cite{Abreu:2019odu,Abreu:2020cwb}. This was investigated in the context of cluster algebras in \cite{Chicherin:2020umh}.

We have already noticed that the first entries involve only five letters since they describe the branching points of the functions. The last entries of the symbol are revealed upon differentiation, see \p{diffpent}. We observe that the last entries of $g^{(2)}_0$ symbol involve 15 linearly independent combinations of 20 letters, and the last entries of $g^{(2)}_1$ symbol involve 14 linearly independent combinations of 19 letters.
In the context of remainder and ratio functions of planar amplitudes in ${\cal N} = 4$ sYM \cite{Drummond:2008vq}, powerful constraints on the last entries of their symbol expressions follow from the $\bar{Q}$-equation \cite{Caron-Huot:2011dec}, which describes the broken super-conformal symmetry of this finite observable. This differential equation provided crucial input for the hexagon bootstrap calculations of multi-loop scattering amplitudes in ${\cal N} = 4$ sYM. The reduced number of letters in the last entries we observe here suggests to us that a $\bar{Q}$-equation for a supersymmetrised $f_5$ could be useful in multiloop calculations. This is currently under investigation \cite{CHT}.

\section{Details of the two-loop calculation and singular limits}
\label{sec:calculationandresult}

\subsection{From amplitude integrands to integrands of $F_n$}

The Wilson loop correlators coming in the definition of $F_n$ \p{FnBos} are not well-defined in $d=4$ dimensions because of the cusp divergences \cite{Drummond:2007au}. However, the ratio $F_n$ is a finite observable in $d=4$ dimensions. Thus, we would like to bypass regularization issues as much as possible staying in four dimensions. More precisely, the aim is to find four-dimensional integrands ${\cal I}^{(L)}$ which represent the loop-corrections $F^{(L)}_n$ in \p{FnPertSer},
\begin{align}
F_n^{(L)} = \frac{1}{(\textup{i}\pi^2)^L L!}  \int d^4 y_{1} \ldots d^4 y_{L}\, {\cal I}^{(L)}_n(x_1,\ldots,x_n;x_0|y_{1},\ldots,y_{L}) \,. \label{FnIntegr}
\end{align}
We assume that the integrands are symmetrized over all $L!$ permutations of the integration points $y_1,\ldots,y_L$. Let us stress that since $F_n^{(L)}$ is finite, the loop integrations in \p{FnIntegr} are well-defined in $d=4$.

We take a shortcut to concise expressions for the planar integrand ${\cal I}^{(L)}$ employing results about integrands of the scattering amplitude in ${\cal N} = 4$ sYM available in the literature. Indeed, the planar MHV scattering amplitude is equivalent (up to regularization scheme) to the vacuum expectation value of the null polygonal Wilson loop at weak as well as at strong coupling (see \cite{Alday:2008yw} for a review),
\begin{align}
\vev{W_n} \sim M_n \equiv \frac{A^{\rm MHV}_n}{A^{\rm MHV}_{\rm tree}}   \,, \label{M=WL}
\end{align}
provided that coordinates (i.e. dual momenta) of the Wilson loop cusps and momenta of scattered particles are related as in \p{xp}. One of the possible ways to make the duality in \p{M=WL} between two divergent at weak coupling objects exact is to consider their integrands. Namely, properly defined four-dimensional $L$-loop integrand of the MHV amplitude and of the Wilson loop are identical \cite{Alday:2010zy,Eden:2010zz,Adamo:2011dq}. 

Following this path, we consider the perturbative expansion
of the planar $n$-particle MHV amplitude
\begin{align}
M_n = 1 + \sum_{L \geq 1} (g^2)^L M_n^{(L)} \label{Mn}
\end{align}
where $M_n^{(L)}$ is formally represented by the planar $L$-loop integrand $I^{(L)}_n$ written in the dual momenta \p{xp},
\begin{align}
M_n^{(L)} = \frac{1}{(\textup{i}\pi^\frac{d}{2})^L L!} \int d^d y_{1} \ldots d^d y_{L} \, I^{(L)}_n(x_1,\ldots,x_n|y_{1} \ldots y_{L}) \,. \label{Mintegr}
\end{align}
The four-dimensional amplitude integrands are very well studied \cite{Arkani-Hamed:2012zlh}. They can be constructed recursively via the loop BCFW \cite{Arkani-Hamed:2010zjl}, and via the twistor space Feynman rules \cite{Mason:2010yk}. They allow for geometric description as dlog forms on the Amplituhedron \cite{Arkani-Hamed:2013jha}. They have also been constructed in manifestly local form \cite{Arkani-Hamed:2010pyv} to high loop orders via from soft and collinear consistency relations \cite{Bourjaily:2011hi}, via a connection to correlation
functions \cite{Eden:2011we,Eden:2012tu}, and via generalized unitarity.

A systematic way to define the Wilson loop integrand---and via the duality $M_n \sim \vev{W_n}$ \p{M=WL} the integrand of the planar MHV amplitude---avoiding ill-defined four-dimensional loop integrations \p{Mintegr} relies on the Lagrangian insertion formula. According to it, differentiation of a correlation function in the coupling results in insertion of the Lagrangian,
\begin{align}
g^2 \pa_{g^2} \vev{W_n} = -\textup{i}\int d^d x_0 \, \vev{W_n {\cal L}(x_0)} \,. \label{LagrIns}
\end{align}
Expanding both sides of the previous relation in the coupling, we find that the correlator $\vev{W_n {\cal L}}$ evaluated at the lowest perturbative level equals to the one-loop integrand of $\vev{W_n}$. Iterating this procedure, one finds that the $L$-loop integrand $I_n^{(L)}$ of the Wilson loop equals to the correlation function of the Wilson loop $W_n$ with $L$ Lagrangians insertions evaluated at the lowest perturbative order. The latter is a finite rational function in four dimensions.

Similarly, the insertion formula \p{LagrIns} implies that the finite ratio $F_n$ \p{FnBos} can be considered as the integrand of the logarithm of the Wilson loop,
\begin{align}
g^2 \pa_{g^2} \log\vev{W_n}  = \int \frac{d^d x_0}{\textup{i} \pi^\frac{d}{2}} \, F_n(x_1,\ldots,x_n;x_0) \,. \label{logWn}
\end{align}
Namely, $F_n$ is a finite function, and only integration over $x_0$ in \p{logWn} results in cusp divergences of $\log\vev{W_n}$. In the dimensional regularization $d=4-2\ep$, they manifest themselves as poles $1/\ep^2$.  On the contrary, each of the dimensionally regularized loop integration in $M_n^{(L)}$ \p{Mintegr} brings in $1/\ep^2$ pole. 

Eq. \p{logWn} and the duality $M_n \sim \vev{W_n}$ enables us to relate the planar integrands ${\cal I}^{(L)}_n$ of $F_n^{(L)}$ from \p{FnIntegr} with planar integrands $I^{(L)}_n$ of the MHV amplitude $M_n$ from \p{Mintegr}.
We expand $\log M_n$ in the coupling, see \p{Mn}, 
\begin{align}
g^2 \pa_{g^2} \log\vev{W_n} = & g^2 M_n^{(1)} + g^4 \left[ 2 M^{(2)}_n - \left(M^{(1)}_n\right)^2 \right]  \notag\\ 
& + g^6 \left[ 3 M^{(3)}_n - 3 M^{(1)}_n M^{(2)}_n + \left(M^{(1)}_n\right)^3  \right] + {\cal O}\left(g^8\right),
\end{align}
and we match the integrands \p{FnIntegr} and \p{Mintegr} on both sides of the previous equation,
\begin{align}
& {\cal I}^{(0)}(x_0) = I^{(1)}(x_0)  \,,\label{Fintegr0loop} \\
& {\cal I}^{(1)}(x_0|y_{1}) = I^{(2)}(x_0,y_{1}) - I^{(1)}(x_0) I^{(1)}(y_{1})    \,,\\
& {\cal I}^{(2)}(x_0|y_{1},y_2) = I^{(3)}(x_0,y_{1},y_2) - I^{(1)}(x_0)I^{(2)}(y_1,y_2) - I^{(1)}(y_1) I^{(2)}(x_0,y_2) \notag\\
& \hspace{3cm} - I^{(1)}(y_2) I^{(2)}(x_0,y_1) + 2 I^{(1)}(x_0)I^{(1)}(y_1)I^{(2)}(y_2) \,, \label{Fintegr2loop}
\end{align}
where we omit dependence of the integrands on the external kinematics $x_1,\ldots,x_n$ for the sake of brevity.
Let us note that the integrands ${\cal I}^{(L)}$ of $F_n^{(L)}$ are symmetric in permutations of $x_0,y_1,\ldots,y_L$, namely $x_0$ appears in the integrand on the same footing with the integration points $y_1,\ldots,y_L$. This is not surprising since ${\cal I}^{(L)}$ is the $(L+1)$-loop integrand of $\log \vev{W_n}$ with integration over dual momenta $x_0,y_1,\ldots,y_L$.

We have already noticed that loop integrations in $F_n$ \p{FnIntegr} are well defined in four dimensions in contrast with ill-defined loop integrations in $M_n$ \p{Mintegr}. This property can be observed at the level of the integrand as well. Namely, the integrands ${\cal I}^{(L)}$ are less singular than the amplitude integrands $I^{(L)}$. The integrand \p{FnIntegr} has the following form with local poles
\begin{align}
{\cal I}^{(L)}_n(x_0|y_1,\ldots,y_L) := \frac{N^{(L)}_n(x_1,\ldots,x_n;x_0|y_1,\ldots,y_L)}{ \prod\limits_{\substack{ 1\leq i<j\leq n\\ 1 <|i-j|<n-1}} x_{ij}^2 \cdot \prod\limits_{i=1}^{n} x_{i0}^2 \cdot \prod\limits_{i=1}^{n}\prod\limits_{j=1}^{L}(x_{i}-y_j)^2}    \label{FnIntegrLoc}
\end{align}
where the numerator $N_n^{(L)}$ is a polynomial in all dual momenta. The amplitude integrand $I^{(L)}$ has an analogous local form, but its numerator is different. The numerator $N^{(L)}$ of ${\cal I}^{(L)}$ in \p{FnIntegrLoc} suppresses potential divergences of the loop integrations.
The cusp divergences of the Wilson loop can be attributed to the regions of loop integrations where one (or several) of the loop variables $y_j$ approaches any light-like edge $[x_i;x_{i+1}]$. The numerator of the integrand ${\cal I}^{(L)}$  vanishes in this regime,
\begin{align}
N^{(L)}\bigl(x_0|y_1,\ldots,y_j = \alpha x_i + (1-\alpha) x_{i+1},\ldots,y_L \bigr) = 0 \,,\quad
0\leq \alpha \leq 1 \,. \label{Nzero}
\end{align}
Because of the permutation symmetry of $x_0,y_1,\ldots,y_L$, the numerator $N^{(L)}$ also vanishes if $x_0$ belongs to any light-like edge,
$x_0 \in  [x_{i};x_{i+1}]$. This property of the numerator serves as a helpful cross-check of the integrand.

In the following, we consider the five-particle case $n=5$. Thus, in order to construct integrands ${\cal I}_5^{(L)}$ according to \p{Fintegr0loop}--\p{Fintegr2loop}, the five-particle planar MHV amplitude integrands are required. The latter have been calculated in \cite{Ambrosio:2013pba} up to the six-loop order with the help of the amplitude/correlator duality \cite{Eden:2010ce,Eden:2010zz,Eden:2011yp}. They are explicitly dual conformal functions with local poles. Since we aim at the two-loop calculation of $F_5$, we use the MHV amplitude integrands of \cite{Ambrosio:2013pba} up to the three-loop order, see \p{Fintegr2loop}. For example,
the one-loop integrand of the five-particle MHV amplitude, see \p{Fintegr0loop},
\begin{align}
{\cal I}^{(0)}_5(x_0) = & - \frac{1}{2} \biggl[ \frac{x_{24}^2 x_{35}^2}{ x_{20}^2 x_{30}^2 x_{40}^2 x_{50}^2} + \frac{x_{14}^2 x_{35}^2}{ x_{10}^2 x_{30}^2 x_{40}^2 x_{50}^2} + \frac{x_{14}^2 x_{25}^2 }{x_{10}^2 x_{20}^2 x_{40}^2 x_{50}^2} + \frac{x_{13}^2 x_{25}^2 }{x_{10}^2 x_{20}^2 x_{30}^2 x_{50}^2} \notag\\ 
&+ \frac{x_{13}^2 x_{24}^2}{x_{10}^2 x_{20}^2 x_{30}^2 x_{40}^2}  + \frac{\ep(x_1,x_2,x_3,x_4,x_5,x_0)}{x_{10}^2 x_{20}^2 x_{30}^2 x_{40}^2 x_{50}^2} \biggr] \label{I0integr}
\end{align}
contains only local poles, and its dual conformal invariance (with weight $+4$ at point $x_0$) is explicit. The integrand contains the parity-odd term which is proportional to the totally antisymmetric tensor
\begin{align}\label{defeps012345}
 \ep(x_1,x_2,x_3,x_4,x_5,x_0) := -\frac{i}{6} \sum_{\sigma \in {\cal S}_6 } (-1)^\sigma x_{\sigma_1}^2 \, \ep(x_{\sigma_2},x_{\sigma_3},x_{\sigma_4},x_{\sigma_5})\,,
\end{align}
where $\ep(a,b,c,d)=\ep_{\mu \nu \rho \sigma} a^\mu b^\nu c^\rho d^\sigma$, $(-1)^\sigma$ stands for the signature of $\sigma$, and where the summation runs over all permutations of $\{ 1,2,3,4,5,0\}$. It is dual-conformally covariant with weight $-1$ with respect to each of the points \cite{Ambrosio:2013pba}. 

The higher-loop MHV amplitude integrands $I^{(L)}_5$ of \cite{Ambrosio:2013pba}, and consequently ${\cal I}^{(L)}_5$, have a form similar to \p{I0integr}. They also involve
the parity-odd $\ep(x_1,x_2,x_3,x_4,x_5,y_k)$ with $k=1,\ldots,L$. All totally antisymmetric parity-odd tensors can be reduced to a single one and distances squared $x_{ij}^2$ due to the following identity,
\begin{align}
\ep(x_1,x_2,x_3,x_4,x_5,x_0)\,\ep(x_1,x_2,x_3,x_4,x_5,x_{0'}) = - \det\left(x_{ij}^2\bigr|^{i=0,1,\ldots,5}_{j=0',1,\ldots,5}\right). \label{epep}
\end{align}

Summarizing, we have at our disposal four-dimensional $(6+L)$-point integrands ${\cal I}^{(L)}_5(x_0)$ which are explicitly dual-conformal and contain only local poles.

\subsection{Integrands in the conformal frame}

\begin{figure}
    \centering
    \includegraphics[height=4.5cm]{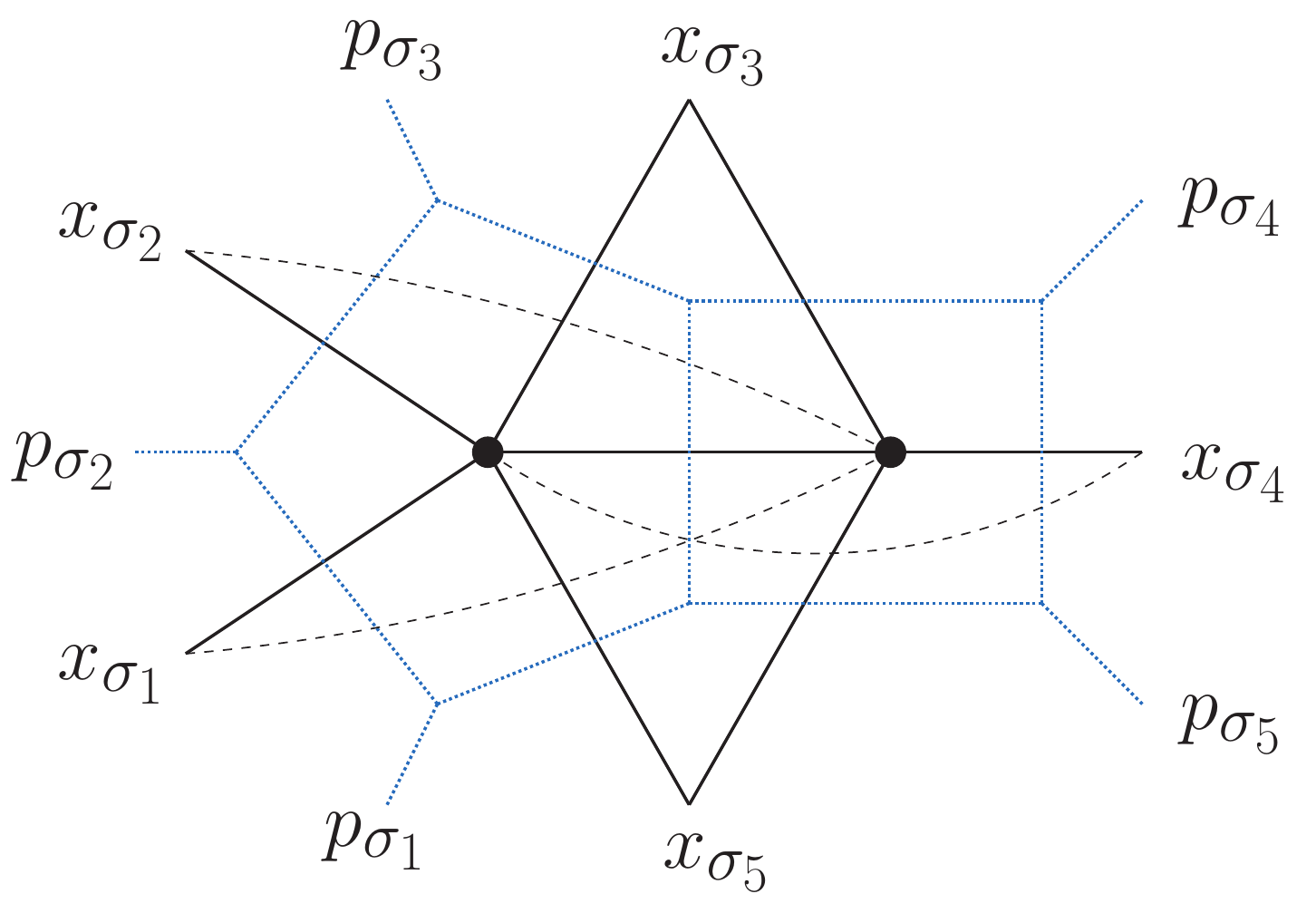}\qquad\qquad
    \includegraphics[height=4.5cm]{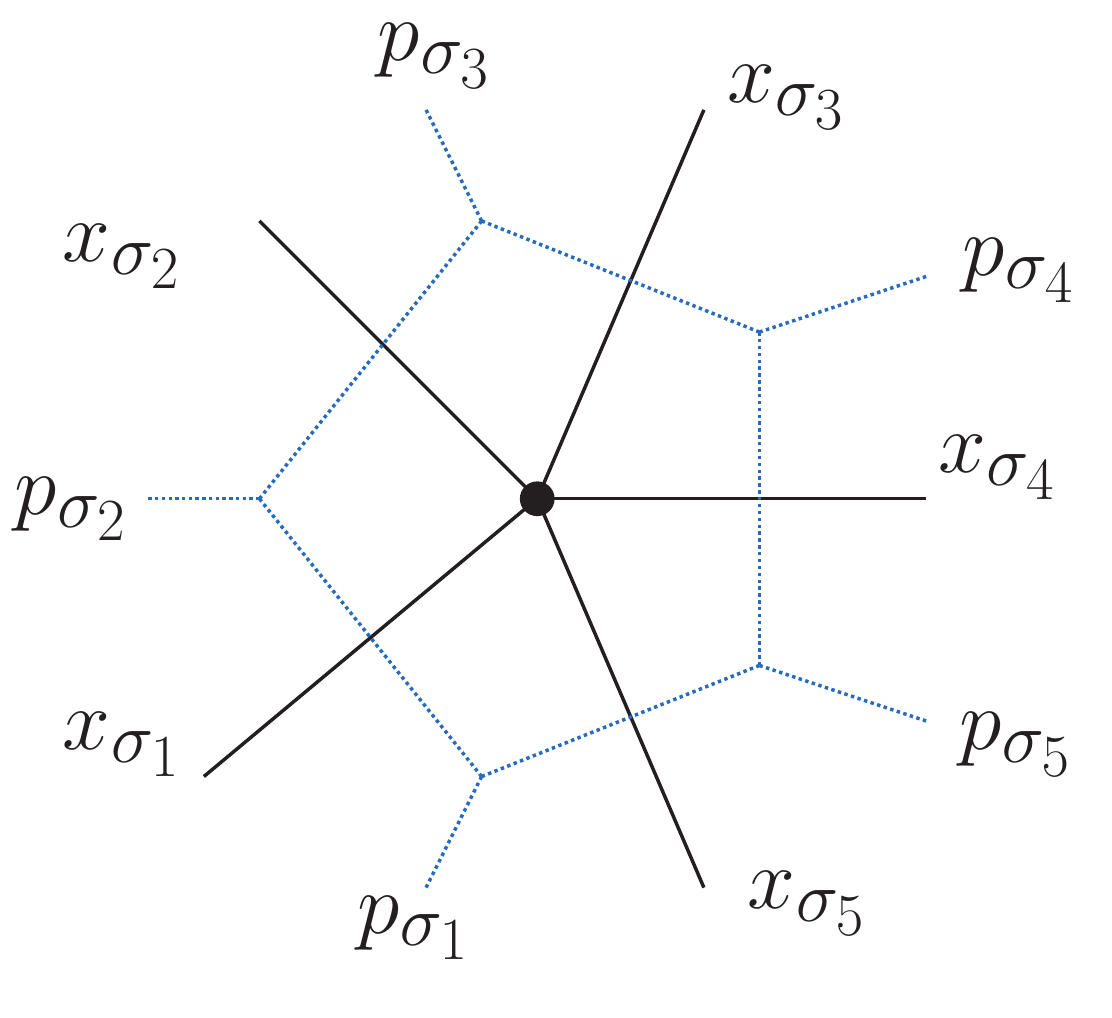}
    \caption{The two-loop pentabox topology (l.h.s.) and the one-loop pentagon topology (r.h.s.) in momentum space (drawn by blue dotted lines) and their dual momentum representation (solid black lines). Integration vertices of the dual graph are denoted by black blobs. Dashed black lines denote numerators. Labels $\sigma_1,\sigma_2,\ldots,\sigma_5$ of the external momenta and the dual momenta are cyclically ordered, such that $x_{\sigma_i}-x_{\sigma_{i-1}} = p_{\sigma_{i}}$.}
    \label{fig:pentabox}
\end{figure}

The integrands ${\cal I}^{(L)}_5(x_0)$ are functions of $6+L$ points. In the frame $x_0 \to \infty$, the dual-conformal integrand simplifies to a Poincar\'{e}-invariant function of $5+L$ points,
\begin{align}
\iota_5^{(L)}(x_1,\ldots,x_5|y_1,\ldots,y_L) := \lim_{x_0 \to \infty} (x_0^2)^4 \, {\cal I}^{(L)}_5(x_1,\ldots,x_5;x_0|y_1,\ldots,y_L) \,.    \label{IntNox0}
\end{align}
For example, the integrand  \p{I0integr} with $L=0$ is the Born-level $f_5$, see \p{eq:resultf5tree}, 
\begin{align}
\iota_5^{(0)}(x_1,\ldots,x_5) = f_5^{(0)} = r_{5,0} \,.    
\end{align}
The parity-odd totally antisymmetric tensor from \p{defeps012345} simplifies to $\ep_5$ \p{e5} in the frame $x_0 \to \infty$,
\begin{align}
\lim_{x_0 \to \infty} \frac{1}{x_0^2}  \ep(x_1,x_2,x_3,x_4,x_5,x_0)   = \ep_5 \,.
\end{align}
In view of the previous relation and \p{epep}, we find that all parity-odd terms of the integrand are proportional to $\ep_5$ in the frame $x_0 \to \infty$. Thus, we decompose the integrand into parity-even and parity-odd terms as follows,
\begin{align}
\iota_5^{(L)} = \iota_5^{(L)+} +\eps_5 \, \iota_5^{(L)-}     
\end{align}
where $\iota_5^{(L)\pm}$ are rational functions of $x_{ij}^2$, $y_{kl}^2$, $(x_i-y_k)^2$ with $k,l=1,\ldots,L$ and $i,j=1,\ldots,5$. They contain only simple local poles in loop variables, namely $y_{kl}^2$ and $(x_i-y_k)^2$ in the denominators. Along with simple local poles $x_{ij}^2$ in the external momenta they also contain spurious nonlocal poles $\Delta$ and $\Delta^2$. The latter cancel out among each other in the final result.

\subsection{From Feynman loop integrals to pentagon functions}

Having the five-particle integrands at our disposal, we proceed to loop integrations. Both parity components of the integrand, $\iota_{5}^{(L)\pm}$, can be integrated in four dimensions. However, individual terms of the integrands would result into divergent loop integrations. In order to implement loop integrations of $\iota_{5}^{(L)\pm}$ term by term, we introduce dimensional regularization $d= 4 - 2\eps$. We are mostly concerned with the two-loop calculation, i.e. $L=2$,
\begin{align}
\frac{1}{2 \left( \textup{i}\pi^\frac{d}{2}\right)^2}\int d^d y_1 d^d y_2 \, \iota_5^{(2)\pm}(x_1,\ldots,x_5|y_1,y_2) \,. \label{inti5}
\end{align}
Each term of the integrand $\iota_{5}^{(2)\pm}$ results into a loop integral which is either a genuine two-loop integral of the pentabox topology or a product of one-loop pentagon topologies, see Fig.~\ref{fig:pentabox}. We employ the amplitude terminology naming the topologies which refers to the momentum representation. Each of the loop integrals is a cyclic permutation of planar topologies, i.e. labels $\sigma_1,\ldots,\sigma_5$ of the external momenta (and dual momenta) in Fig.~\ref{fig:pentabox} are a cyclic permutation of $1,2,\ldots,5$.

The Feynman integrals of the pentabox topology are as follows (see l.h.s. of Fig.~\ref{fig:pentabox})
\begin{align}
\frac{1}{\left(\textup{i}\pi^\frac{d}{2}\right)^2}\int \frac{d^d y_1 d^d y_2}{\prod\limits_{i=1}^{5}(y_1 -x_{\sigma_i})^{2\alpha_i} \; \prod\limits_{i=1}^{5}(y_2 -x_{\sigma_i})^{2\beta_i} \; (y_1-y_2)^{2}} \label{PBscal}
\end{align}
where integer propagator powers $\alpha_i,\beta_i$ satisfy $\alpha_i,\beta_i \leq 1$ and  $\alpha_4,\beta_1,\beta_2\leq 0$; namely, they contain up to eight simple propagators and three irreducible scalar products forming the numerator. The product of two pentagon topologies also contain only simple propagators, i.e. $\alpha_i,\beta_i \leq 1$, 
\begin{align}
\int \frac{d^d y_1}{\textup{i}\pi^\frac{d}{2}} \frac{1}{\prod\limits_{i=1}^{5}(y_1 -x_{\sigma_i})^{2\alpha_i}} 
 \int \frac{d^d y_2}{\textup{i}\pi^\frac{d}{2}} \frac{1}{ \prod\limits_{i=1}^{5}(y_2 -x_{\sigma_i})^{2\beta_i} } \,. \label{PPscal}
\end{align}
The two-loop pentabox integrals and the one-loop pentagon integrals are expressible in the basis of 61 and 11 master integrals, respectively, due to the IBP relations \cite{Chetyrkin:1981qh}. The basis of pure master integrals is a distinguished one \cite{Henn:2013pwa}, since the pure master integrals evaluate to a particularly compact expressions. For the planar pentabox topology, it was constructed in \cite{Gehrmann:2015bfy,Gehrmann:2018yef}. We implement IBP reductions of the scalar Feynman integrals \p{PBscal} and \p{PPscal} to the pure basis by means of \texttt{FIRE6} \cite{Smirnov:2019qkx}.

The pure master integrals are Laurent series in the dimensional regularization parameter $\eps$. The coefficients of the series are polynomials (over rational numbers) in the pentagon functions (see Sect.~\ref{sect:pentfun}) and several transcendental constants \cite{Gehrmann:2018yef}. Upon assigning transcendental weight $-1$ to $\eps$, the Laurent series expansion of the pure master integrals is of homogeneous transcendental weight.

The pure master integrals are soft/collinear divergent that is manifested as poles in the dimensional regularization parameter $\eps$. Individual terms of \p{inti5} contain poles up to order $1/\eps^4$. The poles cancel out in the sum of all terms in agreement with finiteness of the quantity $f_5^{(2)}$ in four dimensions, and we can put $\eps =0$. 

Finally, we find $f_5^{(2)}$ has six leading singularities in agreement with \p{fL5twoLoop}. The accompanying them six pure functions $g^{(2)}_0$ and $g^{(2)}_i$ are polynomials in the planar pentagon functions and transcendental constants $\pi,\zeta_3$ (as well as $\phi^{(4)}_0$ and $\phi^{(4)}_1$) of homogeneous transcendental weight four (see Sect.~\ref{sec:res2loop} and ancillary files for the explicit expressions).

\subsection{Validity checks and singular limits}

\begin{figure}
\begin{align*}
    \begin{array}{c}\includegraphics[height=2.2cm]{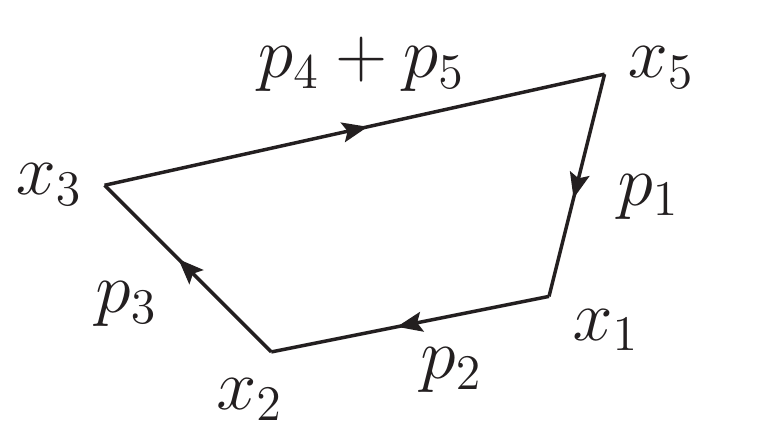}
    \end{array}
    \;\xleftarrow{p_4||p_5}\;
    \begin{array}{c}
    \includegraphics[height=2.7cm]{5WLcontour.eps}\end{array} \;\xrightarrow{p_5\to 0}\;
    \begin{array}{c}\includegraphics[height=2.2cm]{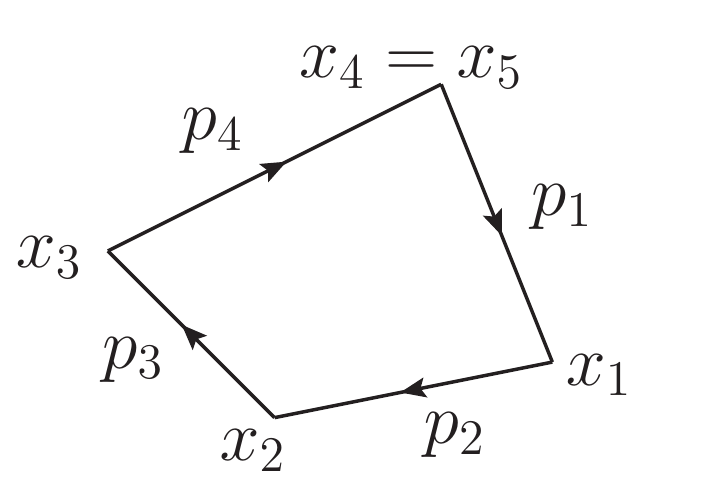}
    \end{array}
\end{align*}  
    \caption{Soft $p_5 \to 0$ and collinear $p_4||p_5$ limit of the five-cusp Wilson loop contour.}
    \label{fig:5ptcontourLimit}
\end{figure}

Scattering amplitudes and Wilson loops are known to simplify in a number of singular regimes. The soft, collinear, and (multi)-Regge limits are of special interest for amplitudes, since they have physical relevance. In the soft and collinear limits a five-particle amplitude reduces to a four-particle amplitude. In these limits, the five-cusp Wilson loop contour turns into a four-cusp contour, see Fig.~\ref{fig:5ptcontourLimit}. We would like to study these limits of the five-particle perturbative data $f_5^{(L)}$. Previously studied four-particle $f_4$ does not allow for undertaking these limits. We expect that the finite observable $f_5$ smoothly reduces in the soft and collinear limits to the four-particle $f_4$ in analogy with the finite remainder of the MHV amplitude in ${\cal N}=4$ sYM.

Indeed, in the soft limit with one of the particle momenta vanishing, $f_5$ reduces to its four-particle counterpart $f_4$, 
\begin{align}
f_{5}^{(L)}(p_1,\ldots,p_5) \to f_4^{(L)}(p_1,\ldots,p_4)   \qquad \text{at} \quad p_5 \to 0 \,. \label{softlim}
\end{align}
We verify \p{softlim} for the available perturbative data $L=0,1,2$. We introduce a parametrization of the kinematics in order to approach smoothly the soft limit preserving the momentum conservation.
The following leading singularities take a very simple form in the soft limit $p_5 \to 0$,
\begin{align}
r_{5,2},\,r_{5,3} \to 0\,,\quad
r_{5,0},\,r_{5,5} \to - s t
\end{align}
where $s,\,t$ \p{st4pt} are Mandelstam invariants of the four-particle kinematics, while $r_{5,1}$ and $r_{5,4}$ are two linearly independent functions which depend on the direction in which the soft limit $p_5 \to 0$ is approached. The accompanying pure functions vanish, and
\begin{align}
g^{(L)}_1,\, g^{(L)}_4 \to 0
\end{align}
so the soft limit takes the following form,
\begin{align}
f^{(L)}_5 \to -s t \lim_{p_5 \to 0}\left( g^{(L)}_0 + g^{(L)}_5 \right) = s t \, G_4^{(L)}(z).
\end{align}
We checked that it matches the four-particle perturbative data \p{Fweak0}--\p{Fweak2}. In the $L=2$ case we checked the statement at the symbol level. 

Similarly, in the limit of collinear $p_4$ and $p_5$, the five-particle $f_5$ smoothly reduces to $f_4$,
\begin{align}
f_{5}^{(L)}(p_1,p_2,p_3,p_4,p_5) \to f_4^{(L)}(p_1,p_2,p_3,P)   \qquad \text{at} \quad p_4 \to x P ,\; p_5 \to (1-x) P \label{collf5}
\end{align}
where $x$ is the splitting parameter and $p_4 + p_5 = P$. The leading singularities are as follows in the collinear limit
\begin{align}
r_{5,1},\,r_{5,2},\,r_{5,3} \to 0\,,\quad
r_{5,0},\,r_{5,4},\,r_{5,5} \to - s t \,,
\end{align}
and the relation between the five-particle and four-particle \p{Fweak0}--\p{Fweak2} perturbative data is as follows
\begin{align}
f_5^{(L)} \to -s t \lim_{p_4||p_5}\left( g^{(L)}_0 + g^{(L)}_4 + g^{(L)}_5 \right) = s t G_4^{(L)}(z)   \,.
\end{align}
In the $L=2$ case we checked the statement at the symbol level.

The leading singularities are divergent if a Mandelstam variable $s_{ij}$ with nonadjacent $i$ and $j$ vanishes, namely $r_{i,5}$ is singular at $s_{i-1\,i+1} = 0$ with $i=1,\ldots,5$. This singularity is suppressed by the pure function $g^{(L)}_{i}$, namely
\begin{align}
g^{(L)}_{i}({\bf u})\Bigr|_{s_{i-1\,i+1} = 0} = 0 \,, \label{spurSing}   
\end{align}
so their product $r_{5,i} \,g^{(L)}_{i}$ is finite at this kinematic point. We explicitly checked this statement for the perturbative data $L=1,2$ (at the symbol level for $L=2$). Let us note that a particular case of the spurious singularity $s_{i-1\,i+1} = 0$ is a kinematical configuration with collinear nonadjacent momenta $p_{i-1}||p_{i+1}$.

Finally, let us recall that the perturbative calculation of $f_5^{(L)}$ with $L=1,2$ involves nontrivial cancellations of soft/collinear poles among the contributing five-particle Feynman integrals, see \p{inti5}. This is a self-consistency check of our calculation. Moreover, the result turned out to be consistent with the maximal transcendentality principle \cite{Kotikov:2002ab} that is common for observables in ${\cal N}=4$ sYM.

\subsection{Multi-Regge limit}
\label{sec:MRK}

\begin{figure}
\begin{center}
\includegraphics[height=3.5cm]{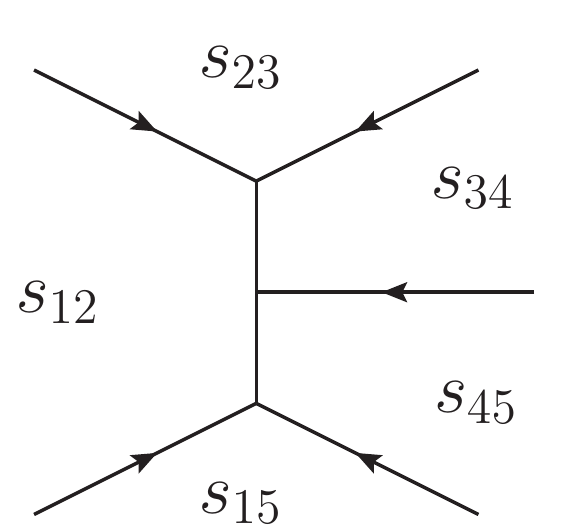}    
\end{center}
    \caption{Five-particle multi-Regge kinematics.}
    \label{fig:5ptMRK}
\end{figure}

The observable $f_{5}$ depends on the same set of variables as five-particle scattering amplitudes, and is closely related to scattering amplitudes. 
Scattering amplitudes are known to simplify and exhibit a lot of structure in a number of asymptotic regimes, such as soft, collinear and high-energy (Regge) limits. 
Therefore it is interesting to study $f_5$ in such limits.
We have already considered the soft and collinear limits of $f_5$. 
Let us investigate high-energy scattering next. Five-particle amplitudes are known to simplify in the multi-Regge kinematics \cite{Kuraev:1976ge,DelDuca:1995hf}. In this regime, the final-state particles are strongly ordered in rapidity and have comparable transverse momenta.

We consider the scattering channel $12 \to 345$. 
The multi-Regge limit is reached via the following parametrization of the kinematics \p{sadj}, together with the limit $\delta \to 0$ (see Fig.~\ref{fig:5ptMRK}),
\begin{align}
& s_{12} =  \frac{ s_1 s_2 }{\kappa\delta^2} \,,\quad  s_{23} = -z_1 z_2 \kappa \,, \quad  s_{34} = \frac{s_1}{\delta} \,, \notag \\  & s_{45} = \frac{s_2}{\delta} \,,\quad  s_{15} = -(1 - z_1) (1 - z_2)  \kappa \,. \label{eq:defMultiReggelimit}
\end{align}
The parity-odd tensor \p{e5} is as follows
\begin{align}
\ep_5 = \frac{s_1 s_2}{\delta^2} (z_1- z_2) + {\cal O}(1/\delta)\,.
\end{align}
In the physical scattering regime $12 \to 345$, we have $(z_1)^* = z_2$ and $s_1,s_2, \kappa > 0$. Whereas in the Euclidean region \p{Eucl5}, $z_1,z_2$ are real and independent (either $z_1 >1, \, z_2 <0$ or $z_1<0,\, z_2 >1$) and $s_1,s_2,\kappa<0$. In the following we consider the Euclidean region of the multi-Regge kinematics.

Since the loop correction of $f_5$ are described by the planar pentagon functions, we apply the multi-Regge limit to the $26$-letter alphabet of planar pentagon function (see Sect.~\ref{sect:pentfun}). We find that it factorizes into four independent alphabets \cite{Chicherin:2018yne,Chicherin:2019xeg,Caron-Huot:2020vlo},
\begin{align}\label{eq:ReggeAlphabet}
\{ 
\delta\} \,, \{ \kappa \} \,, 
\{ s_1 \,, s_2\,, s_1-s_2\,, s_1 + s_2 \} \,,
\{ z_1 \,, z_2 \,, 1-z_1 \,, 1-z_2 \,, z_1 - z_2\,, 1-z_1 -z_2 
\}\,.
\end{align}
It tells us that any pentagon function, in the multi-Regge limit (\ref{eq:defMultiReggelimit}), turns into (a sum of) functions that take the following form,
\begin{align}
(\log \delta)^a \times (\log \kappa)^b \times H_s(s_1,s_2) \times H_z(z_1,z_2) \,, \label{functsMRK}
\end{align}
for some non-negative integers $a,b$, and where $H_s$ are Harmonic Polylogarithms (HPL) \cite{Remiddi:1999ew} described by the $s_1,s_2$-dependent alphabet letters given in Eq. (\ref{eq:ReggeAlphabet}) only, and similarly $H_z$ are 2dHPL functions \cite{Gehrmann:2000zt} with the $z_1,z_2$-dependent alphabet letters. In fact, we find that the leading term of the multi-Regge asymptotics takes a much simpler form  than \p{functsMRK} at $L=1,2$. It is given by five logarithmic functions, see \p{glogs}.

The leading singularities $r_{5,0}$ and $r_{5,4}$ are dominant in the multi-Regge asymptotics \p{eq:defMultiReggelimit}. Their leading terms coincide, 
\begin{align}
r_{5,0} ,\, r_{5,4} = - z_1(1-z_2) \frac{s_1 s_2}{\delta^2} + {\cal O}\left(\delta^{-1}\right) ,
\end{align}
while the remaining four leading singularities are subleading in the multi-Regge limit,
\begin{align}
r_{5,1},\, r_{5,2},\, r_{5,3},\, r_{5,5} = {\cal O}\left(\delta^{-1}\right) . 
\end{align}
Thus, it reasonable to normalize the loop corrections of $f_5$ \p{fL5} by the Born-level $f_5^{(0)}$ \p{eq:resultf5tree}. The ratio contains only logarithmic divergences at $\delta \to 0$, 
\begin{align}
\frac{f_5^{(L)}}{f_5^{(0)}} & =   g^{(L)}_0 + g^{(L)}_4 + {\cal O}\left(\delta \log^{2L}(\delta)\right) \notag\\
& = \sum_{k=0}^{2L} \log^{k}(\delta)\,  h_{k}^{(L)}(z_1,z_2,s_1,s_2,\kappa) + {\cal O}\left(\delta \log^{2L}(\delta)\right) , \label{hMRKlim}
\end{align}
where $h_k^{(L)}$ is of transcendental weight $2L-k$. We calculated that the leading term of the multi-Regge asymptotics $h_{k}^{(L)}$ at the one-loop and two-loop order (i.e. $L=1,2$) relying on the approach of \cite{Caron-Huot:2020vlo}. We find that in the Euclidean region they are given by polynomials in logarithms
\begin{align}
&\log\left(\frac{s_1}{\kappa}\right) \,,\; \log\left(\frac{s_2}{\kappa}\right) \,,\;\log(-z_1 z_2) \,,\; \log\left(-(1-z_1)(1-z_2)\right) \label{glogs}
\end{align}
and the transcendental constants $\pi^2$, $\zeta_3$. The explicit expressions are provided in the ancillary files. 

Finally, based on the available perturbative data, we notice that both the logarithm of the four-particle and five-particle observables have a very simple and similar expressions in the Regge and multi-Regge limits, respectively. The leading logarithmic (LL) and next-to-leading logarithmic (NLL) terms of the four-particle observable \p{F4G4} in the high-energy limit $z \to 0$ are as follows, 
\begin{align}
\log\left(\frac{f_4}{g^2 f_4^{(0)}}\right) = & \left(-g^2 + \pi^2 g^4 + {\cal O }\left(g^6\right) \right) {\bf L}^2 +   \left( -4\zeta_3 g^4  + {\cal O }\left(g^6\right) \right){\bf L} + {\cal O }\left({\bf L}^0\right) \, \label{logf4regge}
\end{align}
where ${\bf L} = \log(z)$, whereas the LL and NLL terms of the five-particle observable in the multi-Regge limit $\delta \to 0$,
\begin{align}
\log\left(\frac{f_5}{g^2 f_5^{(0)}}\right) = & \left(-g^2 + \pi^2 g^4 + {\cal O }\left(g^6\right) \right) {\bf L}^2 
+ \Bigl( \left(g^2 -\pi^2 g^4\right) \log\left(\frac{s_1^2 s_2^2}{\kappa^4 z_1 z_2 (1-z_1)(1-z_2)} \right) \notag\\ 
& -4\zeta_3 g^4  + {\cal O }\left(g^6\right) \Bigr){\bf L} + {\cal O }\left({\bf L}^0\right) \label{logf5regge}
\end{align}
with ${\bf L} = \log(\delta^2)$.
In particular, we observe that the leading logarithmic terms in the four-particle \p{logf4regge} and five-particle \p{logf5regge} cases are identical. It would be very interesting to understand the patterns observed in eqs. \p{logf4regge} and \p{logf5regge} better.

\section{Positivity properties of the five-particle observable}
\label{sec:positivity}

\subsection{Positivity hypothesis}

The Amplituhedron provides a geometric construction of the finite loop integrands of $F_n$ associating them to negative geometries that was demonstrated in the four-particle case \cite{Arkani-Hamed:2021iya}. 
A related interesting question is about positivity properties of the integrated quantities. 
In section \ref{f4pt}, we reviewed the known positivity properties found in the four-particle case.
Here we wish to perform a similar analysis at five points. 
As discussed in section \ref{sec:5partkinAmplituhedron}, the Amplituhedron geometry suggests to us to focus on the Euclidean region, with $\eps_5>0$.
Therefore, the hypothesis that we wish to test, using the available perturbative data, is
\begin{align}
f_5^{(L)}\bigr|_{\rm Eucl^{+}} < 0 \; \text{ at even } L\,,\quad {\rm and} \quad 
f_5^{(L)}\bigr|_{\rm Eucl^{+}} > 0 \; \text{ at odd } L \,, \label{f5allloopsign}
\end{align}
where the subscript stands for the Euclidean region, eq. \p{Eucl5}, with $\ep_5 >0$. 

Note that compared to the four-particle case, the five-particle $f_5^{(L)}$ depends on four cross-ratios ${\bf u}$ \p{eq:uvarsmomentum}. Indeed, the five-particle loop corrections $L >0$ involve pentagon functions which are multivariable functions of the polylogarithmic type. This makes the five-particle positivity hypothesis \p{f5allloopsign} even more nontrivial than the four-particle positivity \p{f4ptPos}.

\subsection{Consistency of positivity hypothesis with singular limits}

Let us begin by verifying that the conditions (\ref{f5allloopsign}) are satisfied in singular limits. 
In the soft and collinear limits $f_5^{(L)}$ reduces to $f_4^{(L)}$, see \p{softlim} and \p{collf5}, 
so the five-particle positivity \p{f5ptPos} reduces to the four-particle positivity \p{f4ptPos}. 

Another simplified setting for testing the positivity properties of $f_5$ is provided by the multi-Regge asymptotics, which we have calculated in Sect.~\ref{sec:MRK} at $L\leq 2$. The sign of the ratio \p{hMRKlim} is dominated by the leading logarithmic term \begin{align}
\frac{f_5^{(L)}}{f_5^{(0)}}=   h^{(L)}_{2L} \left(\log \delta \right)^{2L} + {\cal O}\left(\left(\log \delta\right)^{2L-1}\right)
\end{align}
at sufficiently small $\delta$. The one-loop and two-loop data, $h^{(1)}_2 = -4$ and $h^{(2)}_4=8$, agree with the positivity conjecture \p{f5allloopsign}.

Next, we elaborate on testing the conjecture \p{f5allloopsign} in the full kinematic setting. We find it convenient to study separately the sign properties of the leading singularities on the one hand, and of the loop corrections (pentagon functions) on the other hand.

\subsection{Amplituhedron subregions}

\begin{table}[]
    \centering
\begin{tabular}{c|ccccc}
\toprule
   subregion & $s_{25}$ & $s_{13}$ & $s_{24}$ & $s_{35}$ & $s_{14}$   \\ \midrule
    $\mathrm{(A)}$ &  $+$ & $+$ & $+$ & $+$ & $+$  \\
  $\mathrm{(B)}$  &   $+$ & $+$ & $-$ & $+$ & $+$ \\
   $\mathrm{(C)}$ &  $-$ & $+$ & $+$ & $+$ & $-$ \\ 
    \bottomrule
\end{tabular}
\caption{Subregions of the five-particle Euclidean region \p{Eucl5} which are specified by the signs of the nonadjacent bi-particle Mandelstam variables $s_{i\, i+2}$. For subregions $\mathrm{(B)}$ and $\mathrm{(C)}$ there exist four similar regions respectively that are obtained by cyclic permutations.}
\label{tab:snonadj}
\end{table}

Prior to presenting the positivity properties of the five-particle leading singularities we need to have a better understanding of the Mandelstam invariants signs in the Euclidean region \p{Eucl5}. The two-particle Mandelstam invariants with adjacent indices are negative, $s_{i\, i +1} <0$, but $s_{ij}$ with nonadjacent $i$ and $j$, namely $s_{25},s_{13}, s_{24},s_{35},s_{14}$, do not have a definite sign in the Euclidean region \p{Eucl5}. Specifying signs of all $s_{ij}$ we fix a particular subregion of the Euclidean region \p{Eucl5}. The signs of $s_{ij}$ cannot be chosen arbitrarily.
They are correlated in a certain way. We observe that Mandelstam invariants $s_{i\,i+2}$ and $s_{i+2\,i+4}$ cannot be negative simultaneously. Indeed, the five-particle momentum conservation implies
\begin{align}
s_{i\, i+2} + s_{i+1\, i+2} + s_{i+2\,i+3} + s_{i+2\,i+4} = 0\,.
\end{align}
Since $s_{i+1\, i+2}<0$ and $s_{i+2\,i+3} <0$ in the Euclidean region \p{Eucl5} then at least one of $s_{i\,i+2}$ and $s_{i+2\,i+4}$ is positive.
Summarizing, we find 11 subregions of the Euclidean region:
\begin{itemize}
    \item One subregion with all five $s_{25},s_{13}, s_{24},s_{35},s_{14}$ being positive;
    \item Five subregions with one of $s_{i i+2}$ being negative
    and others being positive;
    \item Five subregions with $s_{i\,i+2}$ and $s_{i+1\,i+3}$ being negative and others being positive.
\end{itemize}
Since $f_{5}$ is invariant under dihedral symmetry, it is sufficient to consider the three cases given in Tab.~\ref{tab:snonadj}.

\begin{table}[]
    \centering
\begin{tabular}{c|c|ccccc}
\toprule
   subregion &   $r_{5,0}$ & $r_{5,1}$ & $r_{5,2}$ & $r_{5,3}$ & $r_{5,4}$ & $r_{5,5}$  \\ \midrule
  $\mathrm{(A)}$ &   $-$ & $+$ & $+$ & $+$ & $+$ & $+$ \\
  $\mathrm{(B)}$ &   $-$ & $+$ & $+$ & $-$ & $+$ & $+$ \\
  $\mathrm{(C)}$ &   $-$ & $-$ & $+$ & $+$ & $+$ & $-$ \\ 
    \bottomrule
\end{tabular}
    \caption{Signs of the leading singularities \p{f50} in different kinematic regions.}
    \label{tab:rsign}
\end{table}

In the following, we ignore lower dimensional subregions with one or several $s_{ij}$ vanishing. In the full-dimensional subregions $(\mathrm{A})$, $(\mathrm{B})$, $(\mathrm{C})$ of the Euclidean region either $\ep_5 >0$ or $\ep_5 <0$.  
We will restrict our considerations to the half of the Euclidean region with $\ep_5 > 0$, as motivated in section~\ref{sec:5partkinAmplituhedron}.

\subsection{Sign properties of the leading singularities and one-loop functions}

Let us now study sign properties of the leading singularities in the regions introduced in the previous subsection.
 We can immediately see that $r_{5,0}$ \p{eq:resultf5tree} is negative in the Euclidean region \p{Eucl5} with $\ep_5 >0$. In order to infer the sign of the remaining leading singularities \p{r5i}, we note that
\begin{align}
\tr_-(p_{i} p_{i+1} p_{i+2} p_{i+3}) < 0  \,,\qquad \text{at} \quad \ep_5 >0 \quad\text{and}\quad {\bf s} < 0 \,.
\end{align}
Then the sign of $r_{5,i}$ is correlated with the sign of $s_{i+1\,i+4}$, 
\begin{align}
r_{5,i} \, s_{i+1\,i+4} > 0  \,,\qquad \text{at} \quad \ep_5 >0 \quad\text{and}\quad {\bf s} < 0 \,.
\end{align}
Thus, the leading singularities have a definite sign inside each of the subregions $(\mathrm{A})$, $(\mathrm{B})$, and $(\mathrm{C})$,
as summarized in Tab.~\ref{tab:rsign}.

\begin{table}[]
    \centering
\begin{tabular}{c|c|ccccc}
\toprule
   subregion &  $g^{(1)}_{0}$ & $g^{(1)}_{1}$ & $g^{(1)}_{2}$ & $g^{(1)}_{3}$ & $g^{(1)}_{4}$ & $g^{(1)}_{5}$  \\ \midrule
 $\mathrm{(A)}$ &    $-$ & $+$ & $+$ & $+$ & $+$ & $+$  \\
   $\mathrm{(B)}$ &    $*$ & $+$ & $+$ & $-$ & $+$ & $+$ \\
  $\mathrm{(C)}$ &     $*$ & $-$ & $+$ & $+$ & $+$ & $-$ \\
    \bottomrule
\end{tabular}
    \caption{Signs of the pure one-loop functions \p{g1loop} in different kinematic regions. The symbol $*$ means that our numerical analysis found both positive and negative signs within the region.}
    \label{tab:gsigns}
\end{table}

We study numerically positivity properties of the one-loop $L=1$ pure functions \p{g01loop} and \p{g1loop} in the Euclidean region \p{Eucl5}. 
Numerical evaluations of the one-loop functions, which are given by logarithms and dilogarithms, does not pose any problem. 
They can be rapidly evaluated with high numerical precision. We probed ${\cal O}(10^{9})$ random points in the Euclidean region. 
Our findings are as follows:
\begin{itemize}
\item Tab.~\ref{tab:gsigns} shows the sign patterns of the different one-loop functions that are suggested by the numerical analysis. 
\item  Comparing to Tab.~\ref{tab:rsign}, one notices that the signs of $r_{5,i}$ and $g^{(1)}_i$ with $i=1,\ldots,5$ match. The signs of each of the six terms in eq. \p{fL5} are collected in Tab.~\ref{tab:rgsigns}.
\item Tab.~\ref{tab:rgsigns} shows that one the six terms can take negative values in regions $(\mathrm{B})$ and $(\mathrm{C})$.\footnote{Interestingly, we notice that $g_0^{(1)} >0$ occurs if the kinematic point is close to the boundaries of the Euclidean region, and $g_0^{(1)} <0$ inside the bulk away from the boundaries.\label{ftntSignOneLoop}} However, our numerical evaluations suggest that their sum is always positive, so that $f_5^{(1)}\bigr|_{\rm Eucl^{+}} > 0$. This is in agreement with analysis in ref.~\cite{Gehrmann:2015bfy}.
\end{itemize}
Let us provide more information on how the sign of the loop functions and the leading singularities could conspire.  
The leading singularity $r_{5,i}$ changes sign crossing the boundary $s_{i-1 \,i+1} = 0$ of the Euclidean subregions, and it is singular on the boundary. However, this singularity is spurious for the observable $f_5$. The loop function $g_i^{(L)}$ vanishes on the the boundary $s_{i-1 \,i+1} = 0$ (recall \p{spurSing}) and it changes sign crossing the boundary. Then the product $r_{5,i}\, g_i^{(L)}$ is smooth and has the same sign on both sides of the boundary $s_{i-1 \,i+1} = 0$ (in its vicinity).

It is worth noting that the constraints $\eps_5>0$ is necessary for positivity of $f_5^{(1)}$. In the subregions $(\mathrm{A})$, $(\mathrm{B})$ and $(\mathrm{C})$ of the Euclidean region with $\eps_5<0$, all six leading singularities are negative-valued
\begin{align}
r_{5,0} < 0 \,, \qquad r_{5,i} < 0  \,,\qquad  i=1,2,\ldots,5 \,,\qquad \text{at} \quad \ep_5 <0 \quad\text{and}\quad {\bf s} < 0 \,.  
\end{align}
At the same time, the one-loop functions \p{g01loop}--\p{g1loop} do not depend on the sign of $\ep_5$. We observe that $f_5^{(1)}$ does not have a definite sign at $\eps_5<0$.

\begin{table}[]
    \centering
\begin{tabular}{c|c|ccccc}
\toprule
 subregion &    $r_{5,0}\, g^{(1)}_{0}$ & $r_{5,1}\, g^{(1)}_{1}$ & $r_{5,2}\, g^{(1)}_{2}$ & $r_{5,3}\, g^{(1)}_{3}$ & $r_{5,4}\, g^{(1)}_{4}$ & $r_{5,5}\, g^{(1)}_{5}$   \\ \midrule
$\mathrm{(A)}$ &     $+$ & $+$ & $+$ & $+$ & $+$ & $+$ \\
  $\mathrm{(B)},\,\mathrm{(C)}$ &    $*$ & $+$ & $+$ & $+$ & $+$ & $+$ \\
    \bottomrule
\end{tabular}
    \caption{Signs of the six terms in the one-loop ($L=1$) expression \p{fL5} of $f_5^{(1)}$ in different kinematic regions.}
    \label{tab:rgsigns}
\end{table}

\subsection{Sign properties of the two-loop functions}

Exploring sign properties of the two-loop functions is more intricate, as we need to evaluate weight three and four pentagon functions. In order to do so, we rely on the computer code \cite{Gehrmann:2018yef}, which is designed to evaluate pentagon functions (of transcendental weight up to four) by doing one-dimensional numerical integrations.\footnote{
We also constructed one-fold integral representations directly for the two-loop functions $g_0^{(2)}$ and $g^{(2)}_i$ bypassing the pentagon function basis. 
We implemented these one-fold integrations in a \texttt{Mathematica} code in order to cross-check evaluations of $g_0^{(2)}$ and $g^{(2)}_i$ based on pentagon function evaluations with the code of \cite{Gehrmann:2018yef}.} 

 We evaluated the pentagon functions at ${\cal O}(10^4)$ phase-space points from each of the three subregions $(\mathrm{A})$, $(\mathrm{B})$, $(\mathrm{C})$. We generated random phase-space points with uniform distribution of the five ${\bf s}$ \p{sadj} in the range $(-5;0)$. Moreover, we separately generated another ${\cal O}(10^4)$ random phase-space points with one or several $s_{ij}$ close to zero in each of the subregions $(\mathrm{A})$, $(\mathrm{B})$, $(\mathrm{C})$. These phase-space points are close to the boundaries of Euclidean region and to the spurious surfaces in the Euclidean region. 
 
As compared to the one-loop situation, we find that the two-loop pure functions provide more sign patterns, see Tab.~\ref{tab:rgsigns2loop}.

Combining this with the signs of the leading singularities, the key features we find, based on the numerical samplings above, are:
\begin{itemize}
\item subregion $(\mathrm{A})$: the signs of 
two-loop pure functions are opposite to the signs of the accompanying leading singularities;
\item subregion $(\mathrm{B})$: if $r_{5,i}<0$ then $g^{(2)}_i > 0$, whereas the remaining two-loop pure functions can take positive or negative values. We managed to identify 20 different sign patterns in this case (see the ancillary files);
\item subregion $(\mathrm{C})$: all two-loop pure functions can take positive or negative values. We identified 23 sign patterns in this case (see the ancillary files).
\end{itemize}

Despite of the fact that six terms of $f^{(2)}_5$ \p{fL5} are not necessarily all negative in subregions $(\mathrm{B})$ and $(\mathrm{C})$\footnote{Similarly to the one-loop observation, see footnote~\ref{ftntSignOneLoop}, our numerical experiments suggest that for the phase space points which are not close to the boundaries $s_{i\,i+1}=0$ of the Euclidean region, the sings of the two-loop functions are opposite to those of the leading singularities. Namely the sign patterns $(+|--+--)$ and $(+|+---+)$ in the second line and the third lines of Tab.~\ref{tab:rgsigns2loop}, respectively, are dominant in the bulk of the Euclidean region. The remaining sign patterns of the two-loop functions are observed in vicinity of the boundaries.} (second and third line of Tab.~\ref{tab:rgsigns2loop}), we observe that their sum is always negative in the Euclidean region with $\ep_5 >0$, i.e. $f^{(2)}_5\bigr|_{\rm Eucl^{+}} < 0$. The above analysis underlines how non-trivial this property is. 

In summary, we have provided numerical evidence that the positivity conjecture \p{f5allloopsign}
holds up to $L = 2$,
\begin{align}
f_5^{(0)}\bigr|_{\rm Eucl^{+}} < 0 \,,\quad
f_5^{(1)}\bigr|_{\rm Eucl^{+}} > 0 \,,\quad
f_5^{(2)}\bigr|_{\rm Eucl^{+}} < 0 \,. \label{f5ptPos}
\end{align}

\begin{table}[]
    \centering
\begin{tabular}{c|c|ccccc}
\toprule
   subregion &   $g^{(2)}_{0}$ & $g^{(2)}_{1}$ & $g^{(2)}_{2}$ & $g^{(2)}_{3}$ & $g^{(2)}_{4}$ & $g^{(2)}_{5}$  \\ \midrule
  $\mathrm{(A)}$ &   $+$ & $-$ & $-$ & $-$ & $-$ & $-$ \\
    $\mathrm{(B)}$ &    $*$ & $*$ & $*$ & $+$ & $*$ & $*$  \\
    $\mathrm{(C)}$ &    $*$ & $*$ & $*$ & $*$ & $*$ & $*$  \\
    \bottomrule
\end{tabular}
    \caption{Observed signs of two-loop pure functions in kinematic regions. Correlations among the signs $*$ are shown in more detail in the ancillary files.
    }
    \label{tab:rgsigns2loop}
\end{table}

\section{Prediction for the three-loop all-plus pure Yang-Mills amplitude}\label{sect:all-plus} 

The scattering amplitude of $n$ gluons in the all-plus helicity configuration in the pure Yang-Mills theory on the one hand, and the Lagrangian insertion in the $n$-cusp Wilson loop in ${\cal N} = 4$ sYM on the other hand, are closely related. In reference \cite{Chicherin:2022bov} we conjectured that the maximal transcendentality parts of the two quantities agree in the planar limit.
Let us briefly review this duality in the general, $n$-point case, and then specialize to the five-point amplitude.

The all-plus amplitude vanishes at the tree level. The one-loop $n$-particle color-ordered all-plus partial amplitude $A^{(1)}_{{\rm YM},n}$ is a rational function \cite{Bern:1993qk}. It equals to the Born-level $n$-particle observable $f_n$ \p{fnLimFn} normalised with the Parke-Taylor factor \p{eq:ParkeTaylor},
\begin{align}
 A^{(1)}_{{\rm YM},n} = {\rm PT}_{n} \, f^{(0)}_{n} \,. \label{A1YMnpt}
\end{align}
We have already mentioned the four-particle \p{PT4r40} and five-particle \p{A1YM5pt} instances of this duality relation.

Higher orders of the all-plus amplitude $A_n^{\rm YM}$ perturbative expansion contain divergences. We consider them in the dimensional regularization with $d= 4 - 2\ep$. The leading color two-loop all-plus amplitude is calculated in \cite{Dunbar:2016cxp,Dunbar:2017nfy}. The duality is formulated in terms of the finite remainders ${\cal H}_n^{\rm MHV}$ and ${\cal H}_n^{\rm YM}$ of the four-particle planar leading-color amplitudes. They refer to the
MHV amplitude of ${\cal N} = 4$ sYM and the all-plus YM amplitude, respectively,
\begin{align}
{\cal H}_n^{\rm MHV}  = \frac{A_n^{\rm MHV}}{A_{n,{\rm tree}}^{\rm MHV} {\cal Z}_{\rm IR}^{\rm MHV}} \;,\qquad
{\cal H}_n^{\rm YM}  = \frac{A_n^{\rm YM}}{g^2 A^{(1)}_{{\rm YM},n} {\cal Z}_{\rm IR}^{\rm YM}}  \,. \label{Hampl}  \end{align}

The infrared renormalization constants ${\cal Z}_{\rm IR}$ minimally subtract divergences, namely $\ep$-expansion of $\log {\cal Z}_{\rm IR}$ contains $\ep$-pole terms but it does not contain finite terms,
\begin{align}
{\cal Z}_{\rm IR} = 1 - g^2 \left( \frac{n}{\ep^2} + \frac{1}{\ep} \sum_{i=1}^{n} \log\left( \frac{\mu^2}{-s_{i \, i+1}}\right)\right)+ {\cal O}(g^4)\,.
\end{align}
The maximally transcendental parts of the renormalization constants coincide in both theories \cite{Kotikov:2002ab},
${\cal Z}_{\rm IR}^{\rm MHV} \sim {\cal Z}_{\rm IR}^{\rm YM}$, that we denote with $\sim$. Let us note that $\log {\cal H}_n^{\rm MHV} + \log {\cal Z}^{\rm MHV}_{\rm IR}$ is known at any loop order at $n\leq 5$ due to the ABDK/BDS ansatz \cite{Anastasiou:2003kj,Bern:2005iz}\footnote{Usually the ABDK/BDS ansatz is presented in a form with non-minimal infrared subtraction.}. Then, the conjecture expresses the maximally transcendental part of the finite all-plus amplitude remainder ${\cal H}_n^{\rm YM}$ in the planar leading-color approximation in terms of the finite $n$-particle observables in ${\cal N}=4$ sYM theory,
\begin{align}
\log {\cal H}_n^{\rm YM}  \sim \log\left(\frac{f_n}{g^2 f_n^{(0)}} \right)  +\log{\cal H}_n^{\rm MHV} +{\cal O}(\ep) \,. \label{dualn}
\end{align}
The appearance of the MHV amplitude in the duality relation \p{dualn} is explained by the well-known duality $A_n^{\rm MHV} \sim \vev{W_n}$ between the planar MHV amplitudes and the polygonal Wilson loops (see \cite{Alday:2008yw,Henn:2020omi} for reviews).

Roughly speaking, the conjecture \p{dualn} relates $(L+1)$-loop all-plus amplitude and $L$-loop $f_n$. In the four-particle case $n=4$, the duality takes the form
\begin{align}
\log {\cal H}_4^{\rm YM}  \sim \log\left( 1-\sum_{L \geq 1} g^{2L} G_4^{(L)} \right)  +\log{\cal H}_4^{\rm MHV} +{\cal O}(\ep) \,, \label{dual4}
\end{align}
with $G_4$ defined in eq.~\p{GPertSer}
that is validated in ref.~\cite{Chicherin:2022bov} for the three-loop all-plus amplitude calculated in \cite{Jin:2019nya,Caola:2021izf}.

As compared to the four-particle all-plus amplitude, the perturbative data for the five-particle all-plus amplitude is available only up to the two-loop order \cite{Gehrmann:2015bfy}. Thus, we can use the duality relation \p{dualn} to provide predictions about the three-loop five-particle all-plus amplitude. Indeed, the finite remainder of the five-particle all-plus amplitude defined in \p{Hampl} has the following perturbative expansion,  
\begin{align}
{\cal H}_5^{\rm YM} := 1+ g^2 {\cal H}_{{\rm YM},5}^{(1)} + g^4 {\cal H}_{{\rm YM},5}^{(2)} + {\cal O}(g^6) \label{H5YM}
\end{align}
where ${\cal H}_{{\rm YM},5}^{(L)}$ describe the $(L+1)$-loop corrections of the all-plus amplitude. Expanding the five-particle duality relation \p{dualn} perturbatively in the coupling, we find,
\begin{align}
{\cal H}_{{\rm YM },5}^{(1)} = & \sum_{i=1}^{5} \frac{r_{5,i}}{f_5^{(0)}} g_i^{(1)} - \frac{1}{2}\mathbf{L}_2 + \frac{5\pi^2}{12}\,, \label{HYM51loop} \\
{\cal H}_{{\rm YM },5}^{(2)} = & \sum_{i=1}^{5} \frac{r_{5,i}}{f_5^{(0)}} \left[ g_i^{(2)} + g_i^{(1)} \left(-g_0^{(1)} + \frac{5\pi^2}{12} - \frac{1}{2} \mathbf{L}_2 \right) \right] \notag\\ 
& + g^{(2)}_0 - \frac{1}{2} \left( g_0^{(1)} \right)^2 + \frac{\pi^2}{3} g_0^{(1)} + \zeta_3 \mathbf{L}_1 + \left( \mathbf{L}_2 - \frac{\pi^2}{6} \right)^2 - \frac{\pi^4}{12} \,,  \label{HYM52loop}
\end{align}
where the one-loop pure functions $g^{(1)}_0$ and $g^{(1)}_i$ are given in \p{g01loop} and \p{g1loop}, respectively. Expressions for the two-loop pure functions $g^{(2)}_0$ and $g^{(2)}_i$ in the pentagon function basis are provided in the ancillary files (see discussion around eq.~\p{g2inf}). We also use shorthand notations
\begin{align}
\mathbf{L}_1 := \sum_{i=1}^{5} \log\left( \frac{\mu^2}{s_{i\,i+1}}\right) \,,\qquad
\mathbf{L}_2 := \sum_{i=1}^{5} \log^2\left( \frac{\mu^2}{s_{i\,i+1}}\right) \,.
\end{align}
The expression \p{HYM51loop} is a special case of the $n$-particle two-loop duality relation that we verified in \cite{Chicherin:2022bov}. 
However, eq.~\p{HYM52loop} is new, and it predicts the maximally transcendental part of the five-particle three-loop all-plus planar amplitude. 

It would be extremely interesting to test our conjecture about the maximally transcendental part of the planar five-particle all-plus amplitude by an explicit three-loop Feynman graph calculation. Furthermore, a three-loop calculation of $f_5$ will provide a conjecture for the maximally transcendental part of the planar four-loop five-particle all-plus amplitude.

\section{Discussion and outlook}
\label{sec:outlook}

In this paper, we studied a null pentagonal Wilson loop with a Lagrangian insertion, $f_5$, and computed for the first time its two-loop corrections. The two-loop result is written as a sum of six conformal leading singularities (as conjectured in \cite{Chicherin:2022bov}), each multiplied by a transcendental function of weight four. The functions needed are pentagon functions that are familiar from two-loop planar five-particle scattering amplitudes in QCD. 

We noticed that similarly to appropriately-defined finite parts of scattering amplitudes \cite{Chicherin:2020umh}, the alphabet needed to express the answer does not require the full planar pentagon alphabet. In particular, the letter $W_{31}$ is not needed. Moreover, we observe that the last entries of the weight-four functions we compute depend on a subset of alphabet letters only. We interpret this as a hint that an analogue of the $\bar{Q}$-equation \cite{Caron-Huot:2011dec}, which was instrumental in the bootstrap of scattering amplitudes in ${\mathcal N}=4$ sYM, could play a role here as well, for a supersymmetrised version of $f_5$. This exciting question is currently under investigation \cite{CHT}.

We studied the analytic result in several singular limits, finding consistency with previous four-particle results in the soft and collinear limits, and providing new data in the multi-Regge limit. These limits played and important role in bootstrap approaches to scattering amplitudes in ${\mathcal N}=4$ sYM. We hope that the data we provide can be valuable for similar analyses for the Wilson loops with Lagrangian insertions. Note that the expressions in the limit that we spell out in the paper contain all logarithmically enhanced and finite terms. If required, power-suppressed terms can readily be obtained from the full analytic result that we provided in terms of pentagon functions, along the lines of \cite{Caron-Huot:2020vlo}.

In reference \cite{Arkani-Hamed:2021iya}, it was found that the four-cusp Wilson loop with Lagrangian insertion has definite sign at each loop order $L$ (the overall sign alternates with $L$) in the Euclidean region. We analyzed $f_5$ numerically in the Euclidean region. We focused on the subset with $\eps_5>0$, which is natural from the point of view of the one-loop Amplituhedron. At one loop, special sign patterns of the integrated answer had already been observed in that region \cite{Gehrmann:2015bfy}. In the present paper, we provided numerical evidence that the two-loop correction has a definite sign in that region, despite that fact that individual terms can have alternating signs throughout the region. We take this as strong evidence for a positivity conjecture of the observable $f_5$. 
It would be very interesting to understand how the positivity property follows from properties of the loop integrands. 

Our computation in the present paper used state-of-the-art scattering amplitude and Feynman integral methods for the loop integrations. However, it was shown in \cite{Arkani-Hamed:2021iya} that the observable can be written in terms of manifestly finite, four-dimensional integrals. The new representations are motivated by the Amplituhedron geometry, using a novel decomposition of the observable in terms of certain `negative' geometries. This offers a new perspective and approach to studying the observables.  
It would be very exciting to harness the power of the four-dimensional approach and apply it to the $n$-particle observable \cite{ACHT2022progress}.

\section*{Acknowledgements}

It is a pleasure to thank Nima Arkani-Hamed, Jaroslav Trnka, and Simone Zoia for stimulating discussions and helpful comments on the draft. This project received funding from the European Union’s Horizon 2020 research and innovation programme under ERC grant {\it Novel structures in scattering amplitudes} (grant agreement No 725110). D.C. is supported by the French National Research Agency in the framework of the {\em Investissements d’avenir} program (ANR-15-IDEX-02).

\appendix

\section{Pentagon functions} 
\label{App:pent}

In this Appendix, we provide more details on the pentagon functions, the underlying pentagon alphabet, and the iterated integral representation for the pentagon functions.

The letters of the planar pentagon alphabet $\{ W_{k} \}_{k=1}^{20} \cup \{ W_k \}_{k=26}^{31}$ are naturally organized in several groups by the cyclic symmetry. The first 20 letters are parity-even, and they are given by linear combinations of the Mandelstam variables, 
\begin{align}
& W_1 = s_{12} \;, & \;
W_{1+i} = \tau^i(W_1) \,, \notag \\
& W_6 = s_{34} + s_{45} \;, & \;
W_{6+i} = \tau^i(W_6)  \,,\notag \\
& W_{11} = s_{12} - s_{45}  \;, & \;
W_{11+i} = \tau^i(W_{11}) \,,  \notag \\
& W_{16} = s_{13}  \;, & \;
W_{16+i} = \tau^i(W_{16}) \,, \label{Weven}
\end{align}
with $ i=1,\ldots,4$. The letter $W_{31} = \ep_5$ is parity even and invariant under cyclic shifts $\tau$. The remaining five letters are parity-odd, namely $d\log(W_{26+i}) \to -d\log(W_{26+i})$ under parity conjugation. They have more complicated expressions,
\begin{align}
W_{26} = \frac{\tr_{-}(p_4 p_5 p_1 p_2)}{\tr_{+}(p_4 p_5 p_1 p_2)}
\;,\qquad
W_{26+i} = \tau^i(W_{26})\;,\qquad i=1,\ldots,4 \,. \label{WLettOdd}
\end{align}

Let us note that five of the leading singularities of $f_5$ allow for factorization in the alphabet letters, e.g.
\begin{align}
(r_{5,1})^2 = W_2 (W_3)^3 W_4 \frac{W_{29}}{W_{20}} \,,
\end{align}
and similarly for its cyclic shifts \p{r5isym}.

The pentagon functions $p_a^{(w)}$ are defined as a linear combination of multi-fold iterated integrals over the pentagon alphabet.\footnote{In ref.~\cite{Gehrmann:2018yef}, the planar pentagon functions of the given weight $w$ are grouped into cyclic orbits. Each orbit contains either a single function, which is invariant under cyclic shifts $\tau$ of the particle labels $p_i \to p_{i+1}$, or five pentagon functions which are related among each other by the cyclic shift.
In the latter case, an index $i = 1,\ldots,5$ distinguishes functions from the same cyclic orbit. So, the pentagon functions of \cite{Gehrmann:2018yef} are labeled as follows. Weight-one: $f^{(i)}_{1,1}$; weight-two: $f^{(i)}_{2,1}$; weight-three: $f^{(i)}_{3,\alpha}$ with $\alpha=1,2,3$ and $f_{3,4}$; weight-four: $f^{(i)}_{4,\alpha}$ with $\alpha=1,\ldots,11$ and $f_{4,12}$. The present notations match with those of \cite{Gehrmann:2018yef} as follows: $f_{w,a} = p^{(w)}_{a}$ and $f_{w,a}^{(i)} = p^{(w)}_{a,i}$. We also imply that $p^{(w)}_a$ refers to both $p^{(w)}_a$ and $p^{(w)}_{a,i}$.} 
Despite the fact that the loop corrections in the five-particle kinematics nontrivially depend only on four dimensionless variables, e.g. the ratios \p{eq:uvarsmomentum}, the pentagon functions $p^{(w)}_a = p^{(w)}_a({\bf s})$ are defined as functions of the five Mandelstam variables \p{sadj}. In this way the cyclic symmetry is not broken.

The collection of pentagon functions, supplemented  with transcendental constants $\pi^2$ and $\zeta_3$, is closed upon differentiation \p{diffpent}. Taking into account the transcendental weight of the constants, the differentiation decrements the weight by one unit,
\begin{align}
d p^{(1)}_a =&  \sum_k A^k_{a} \, d\log W_k \,, \label{DEpent1}\\
d p^{(2)}_a = & \sum_{k} A^k_{a,b} \, p^{(1)}_b d\log W_k \,,\\
d p^{(3)}_a = & \sum_{k} \left[ A^k_{a,b}\, p^{(2)}_b + A^k_{a,b,c}\, p^{(1)}_b p^{(1)}_c + A^k_a \,\pi^2 \right] d\log W_k \,,\\
d p^{(4)}_a = & \sum_{k} \Bigl[A^k_{a,b}\, p^{(3)}_b + A^k_{a,b,c} \, p^{(2)}_b p^{(1)}_c + A^{k}_{a,b,c,d}\, p^{(1)}_b p^{(1)}_c p^{(1)}_d  \notag\\ 
& + A'^k_{a,b} \, \pi^2 p^{(1)}_b + A^k_a\, \zeta_3 \Bigr] d\log W_k \,, \label{DEpent4}
\end{align}
where $A^k_{a,b},A^k_{a,b,c},\ldots$ are rational numbers and we omit summation over repeated labels of the pentagon functions.

Differential equations~\p{DEpent1}--\p{DEpent4} can be straightforwardly integrated. In order to do so, one chooses a base point ${\bf s}_0$ in the Euclidean region \p{Eucl5}, e.g.
\begin{align}
{\bf s}_0   \;: \quad  s_{12} = s_{23} = s_{34} = s_{45} = s_{15} = -1 \,,
\end{align}
and connects it with any other point ${\bf s}$ of the Euclidean region by a path $\gamma({\bf s}_0 , {\bf s})$.
Provided pentagon function values at ${\bf s}_0$ are given,\footnote{All planar pentagon functions (up to weight four) of \cite{Gehrmann:2015bfy} are defined to vanish at the Euclidean base point ${\bf s}_0$ except for $p^{(3)}_4$. The latter is given by a weight-3 transcendental constant $p^{(3)}_4({\bf s}_0) = \frac{2}{3} d_{37,3}$ defined in \cite{Gehrmann:2015bfy}.} we evaluate them at ${\bf s}$, e.g.
\begin{align}
p^{(2)}_a({\bf s}) = \sum_{k} A_{a,b}^{k}\, [p^{(1)}_b,W_k]({\bf s})
\end{align} 
where we use shorthand notations 
\begin{align}
[p_a^{(w)},W_k]({\bf s}) := \int\limits_{\gamma({\bf s}_0 , {\bf s})} d\log(W_k) \, p^{(w)}_a \,.
\end{align}
Iterating this procedure, we express $p^{(w)}_a$ as a linear combination of $w'$-fold iterated integrals with $w' \leq w$. For example, the one-loop functions $g^{(1)}_0$ and $g^{(1)}_1$ \p{g1loop} are the following two-fold iterated integrals,
\begin{align}
g^{(1)}_0 =& \left[ \frac{W_1}{W_{3}} , \frac{W_{5}}{W_{2}} \right] + \left[ \frac{W_{5}}{W_{2}} , \frac{W_1}{W_{3}} \right] -\frac{\pi^2}{6}+ {\rm cyclic} \,,\\
g^{(1)}_1=& \left[ \frac{W_{1}}{W_{3}} , \frac{W_{13}}{W_{5}} \right] + \left[ \frac{W_{5}}{W_{3}} , \frac{W_{15}}{W_{1}} \right] +\frac{\pi^2}{6} \,.
\end{align}

If we omit transcendental constants in the iterated integral expressions, we obtain their symbol counterparts. So, the symbol of a weight-$w$ pentagon function is a linear combination (with rational coefficients) of length-$w$ words $[W_{i_1},W_{i_2},\ldots,W_{i_w}]$ formed from the letters of the pentagon alphabet.

The letters of the pentagon alphabet have definite parity, i.e. $\{ W_{i}\}_{i=26}^{30}$ \p{WLettOdd} are parity-odd and $\{ W_{i}\}_{i=1}^{20}$ \p{Weven} and $W_{31}$ are parity even. The parity grading is inherited by the symbols as well, namely a parity-odd word $[W_{i_1},W_{i_2},\ldots,W_{i_w}]$ involves odd number of parity-odd letters. The planar pentagon functions and the corresponding iterated integrals have definite parity as well.\footnote{Among the planar pentagon functions of \cite{Gehrmann:2015bfy} the parity-odd are $p^{(3)}_4, p^{(4)}_{12}$ and $p^{(4)}_{10,i}$ with $i=1,\ldots,5$. We would also need to assign odd parity to the transcendental constant $d_{37,3}$ in order the iterated integrals have definite parity.}

\bibliographystyle{JHEP}
\bibliography{wl5plus1.bib}

\end{document}